**A simple spontaneously active Hebbian learning model: homeostasis of activity and connectivity, and consequences for learning and epileptogenesis**

David Hsu (1), Aonan Tang (2), Murielle Hsu (1), and John M. Beggs (2)

1. Department of Neurology, University of Wisconsin, Madison WI, United States

2. Department of Physics, Indiana University, Bloomington IN, United States

Contact information: David Hsu, Department of Neurology, University of Wisconsin Hospital & Clinics, 600 Highland Av, Madison WI, 53792 USA. Email:

hsu@neurology.wisc.edu

**Abstract**

A spontaneously active neural system that is capable of continual learning should also be capable of homeostasis of both firing rate and connectivity. Experimental evidence suggests that both types of homeostasis exist, and that connectivity is maintained at a state that is optimal for information transmission and storage. This state is referred to as the critical state. We present a simple stochastic computational Hebbian learning model that incorporates both firing rate and critical homeostasis, and we explore its stability and connectivity properties. We also examine the behavior of our model with a simulated seizure and with simulated acute deafferentation. We argue that a neural system that is more highly connected than the critical state (i.e., one that is "supercritical") is epileptogenic. Based on our simulations, we predict that the post-seizural and post-deafferentation states should be supercritical and epileptogenic. Furthermore, interventions that boost spontaneous activity should be protective against epileptogenesis.

**Introduction**

The destabilizing effects of associative (Hebbian) learning on neural systems have long been recognized: favorable synapses become stronger, unfavorable synapses become weaker, and there is a strong drive either towards ever escalating, runaway activity or else global network silence [1-6]. Activity-dependent regulation of synaptic strengths and intrinsic membrane currents have been proposed as mechanisms of maintaining homeostasis of neuronal activity [7]. Because of these homeostatic mechanisms, neuronal circuits can behave similarly even when the cellular level constituents of neuronal behavior are different [8, 9]. A particularly striking example of a synaptic mechanism of homeostasis is multiplicative post-synaptic up-regulation of AMPA receptors as a compensatory mechanism for synaptic blockade [10, 11].

At present, studies of neural systems homeostasis primarily focus on activity or firing rate homeostasis, whereby the firing rate of a system is constrained to approach a target level of activity [4, 5, 12, 13]. However, although firing rate homeostasis can maintain stability of activity, it does not guarantee a neural system that is functionally useful. For instance, a trivial solution for firing rate homeostasis is to cut all neuronal connections, and then to adjust the spontaneous (or intrinsic) firing rate for each neuron until the target firing rate is reached. Such a neural system is incapable of learning or information processing, as every neuron is isolated from every other neuron. Conversely, a connectivity pattern such that activation of one neuron reliably causes activation of all other neurons is equally useless. Neurons that are capable of learning must maintain an intermediate level of connectivity, even in the face of synaptic changes brought about by Hebbian learning and firing rate homeostasis. Therefore, we expect at least one other homeostatic principle, one for homeostasis of connectivity. Such a mechanism would prevent the neural system from entering a state that is either under- or over-connected.

Local field potential measurements have demonstrated evidence for homeostasis of connectivity. In acute cortical slices and in cortical slices cultured on 60 channel microelectrode arrays, activity consists of periods of quiescence broken by bursts of activity of any number of electrodes, which occur in clusters (or "avalanches") of all possible sizes [14-16]. A branching ratio $\sigma$ can be defined as the number of electrodes that are excited after any other single electrode is excited, averaged over time and over all electrodes. This ratio fluctuates about unity for hours at a time [14, 15], with root mean square deviations of about 5-25% (Beggs, unpublished). The condition $\sigma = 1$ represents the critical point. The return to criticality after fluctuations away from criticality represents critical homeostasis.

The critical point so defined has a number of interesting properties [17]. Chief among them are that critical branching optimizes information throughput [14] and maximizes information storage capacity [15, 16]. A related property is that the probability distribution function of avalanche sizes $G_A(n)$ obeys a power law with $G_A(n) \sim n^{-1.5}$ [14]. The power law behavior suggests that the neural system has long range correlations and is able to access the entire repertoire of possible activation states. Such systems are also referred to as being scale-free [18-21]. That the critical point is optimal for information throughput and storage capacity, and that critical systems have scale-free access to the entire range of possible activation patterns, are highly desirable properties for learning and information processing systems. Thus critical homeostasis represents a homeostasis of connectivity optimized for learning and information processing.

That a branching ratio of $\sigma = 1$ is optimal for information storage and processing is intuitively plausible. If $\sigma > 1$, then one might expect loss of information during transmission due to its being "whited out" by accelerating network activity. If $\sigma < 1$, then information is lost due to its being damped out. A branching ratio of $\sigma = 1$ is most likely to preserve signal transmission in a faithful way. Similarly, for information storage, one might think of information as being stored in the patterns of network activation. If $\sigma < 1$, then these patterns are limited to small cluster sizes, while if $\sigma > 1$, then only large clusters can be activated. At $\sigma = 1$, however, it is possible to activate clusters of all sizes, from the smallest sizes up to clusters the size of the entire network. It is at $\sigma = 1$ that one has the largest repertory of patterns of activation, and thus the largest information storage capacity.

The biomolecular mechanism for critical homeostasis is not known. One might conjecture that such a mechanism depends on the concentration of intracellular calcium, on trophic factors such as brain derived neurotrophic factor, on backpropagation of action potentials into the proximal dendritic tree, or on competition for cellular resources [2, 5, 13]. Interneuron interactions are another candidate. In any case, we *postulate* that such a mechanism exists, particularly for brain regions most involved in learning and high order information processing. Real neural systems might conceivably fluctuate about criticality, or might fluctuate about a point not quite at criticality. It may also turn out that components of a neural system may be tuned to a state far from criticality. Nonetheless, at a large enough lengthscale, we hypothesize that any neural system that is capable of continual learning must be capable of some degree of critical homeostasis.

Here we investigate the properties of a simple model that incorporates both firing rate homeostasis and critical homeostasis. Our purpose is not to propose a model that is quantitatively correct in every detail, but to investigate first whether it is possible to construct such a model, and second whether there are algorithmic consequences of imposing both firing rate and critical homeostasis for this particular model. In the discussion section, we compare our model with other recent homeostatic models.

**Methods**

The model consists of a set of nodes labeled by $i = 1$ to $N = 64$, each of which has an activation level $A(i,t)$. The activation level $A(i,t)$ gives the probability that node $i$ "fires" at some point in the time interval $(t - \delta t, t]$, where $\delta t$ is the timestep. Each node represents the local field average over some number of neurons near one microelectrode. Each firing event represents a population spike consisting of the near-simultaneous action potential discharge of a subpopulation of nearby neurons. In practice, the firing events are taken to be local field potential spikes that are more than 3 standard deviations above the background amplitude. We speak of nodal firing events to distinguish communal action potential discharges from, for instance, lower amplitude post-synaptic potentials. For the system of Haldeman and Beggs [16], with microelectrode diameter of 0.02 mm and interelectrode distance of 0.2 mm, the number of neurons contributing to a nodal firing event may be as large as 1000.

The nodal local field potential dynamics that we model is not simply related to the underlying neuronal dynamics. Neuronal dynamics consists of the action of ions, molecules, macromolecular and cellular-level structures, while nodal dynamics consists of the interaction of many principal neurons and local interneurons in one field, with principal neurons and interneurons in another distant field. In general, the timescale of larger scale dynamics is slower than that of smaller scale systems, because of the possibility of the emergence of collective modes involving the mass action of many neurons acting in phase with one another [22]. A similar phenomenon is well-known in solid state physics, where the high frequencies of molecular vibrations and translational motion are transformed, when these molecules are condensed into a solid, into low frequency as well as high frequency bands. The low frequency bands correspond to collective vibrations involving many molecules moving in phase relative to each other, while the high frequency bands correspond to intrinsic intramolecular motions. Thus the possibility that nodal timescales may be different from neuronal timescales should not surprise us.

The theoretical bridging of microscopic with macroscopic behavior is a monumental task, beyond the scope of this paper [23-26]. Nonetheless, we expect nodal dynamics to reflect neuronal dynamics in a qualitative way. Thus if neuronal dynamics exhibits firing rate homeostasis, then nodal firing rates must also exhibit firing rate homeostasis. Similarly, Hebbian learning at the neuronal level must also be reflected in correlations in nodal dynamics. We will also assume that every node is equivalent. By equivalent, we mean that every node has the same steady state firing rate and the same steady state branching ratio. Heterogeneous systems may be generalized from what follows.

Let $F(i;t) = 1$ mean that node i fired at some point in the time interval $(t – \delta t, t]$, and $F(i;t) = 0$ mean that it was quiescent. Let $\tau_o$ be the target average time interval between firings, so that $1/\tau_o$ is the target average firing rate. The value of $F(i;t)$ can only be zero or one, never anything inbetween.  However, if < ... > represents a time average over a time period that is much longer than $\tau_o$, then one expects that $<F(i;t)>$ becomes approximately equal to $<A(i;t)>$ which itself approaches $\delta t/\tau_o$.

Let $P(i,j;t)$ be the conditional probability that firing at node j in the time interval (t – $\delta t$, t] causes firing at node i in the next time interval (t, t + $\delta t$], with $P(i,i;t) = 0$.  We refer to the $P(i,j;t)$'s as the stimulated firing probabilities, or equivalently as the connection strengths.  Let $S(i;t)$ be the probability of spontaneous firing at node i in the time interval (t, t + $\delta t$].  This probability does not depend on whether any other node has recently fired.  The total probability that node i will fire in the time interval (t, t + $\delta t$] is then given by one minus the probability that it will not fire, i.e.,

$$A(i;t + \delta t) = 1 - [1 - S(i;t)] \prod_{j=1}^{N} [1 - P(i,j;t) F(j;t)]. \qquad \text{Eq (1)}$$

Here $\Pi_j$ signifies a product over all j from 1 to the total number of nodes, N.  In deriving this equation, we have assumed that the connection strengths $P(i,j;t)$ are mutually independent, so that the probability of multiple mutually independent processes occurring at the same time is given by the product of the individual probabilities.  Multinodal events are not ignored but are approximated by the product of one- and two-nodal probabilities.  More sophisticated models would have to take into account simultaneous multinodal interdependence, including terms of the form $P(i,j,k;t)$ for simultaneous interactions between three nodes, for instance.  We leave these higher order correction terms for future studies.  They may prove important but are much more complex to analyze.

For convenience, define a relative firing rate, $f(i;t) = <F(i;t)>_a \tau_o/\delta t$, where $<...>_a$ represents a time average over the prior time interval $\tau_a$ with $\tau_a >> \delta t$. We take $\tau_a = \tau_o$ for simplicity. Thus firing rate homeostasis requires $f(i;t)$ to hover about unity, and the fluctuation $\Delta f(i;t) = f(i;t) - 1$ to fluctuate about zero. Physiologically, $\tau_a$ is the timescale on which each node, through regulatory feedback mechanisms, perceives what its time-averaged firing rate is, relative to its target firing rate. This timescale has to be on the timescale of $\tau_o$ or longer, because a given feedback time cannot yield an average firing interval longer than itself. We take $\tau_a = \tau_o$ primarily for demonstration purposes, but also note that this timescale is the shortest timescale at which a node can compare actual to target firing rates. This choice thus optimizes nodal responsiveness to fluctuations away from target nodal behavior. In what follows we found it most challenging to achieve stable solutions with smaller values of $\tau_a$; firing rate and critical homeostasis are easier to achieve for larger values of $\tau_a$, presumably because fluctuations about the steady state are smoothed out.

Nodal connectivity can be defined in terms of an input ratio $\eta(i;t)$:

$$\eta(i;t) = \sum_{j=1}^{N} P(i,j;t). \qquad \text{Eq (2)}$$

Here $\Sigma_j$ signifies a sum over all j from 1 to the total number of nodes, N. Analogously, the branching ratio for node i can be defined as:

$$\sigma(i;t) = \sum_{j=1}^{N} P(j,i;t). \qquad \text{Eq (3)}$$

The input ratio is a measure of excitatory input into each node. The branching ratio is a measure of excitatory output from each node. The input ratio averaged over all nodes is equal to the corresponding average of the branching ratio. We define critical homeostasis in terms of homeostasis of the input ratio rather than of the branching ratio, but either method will work. Our choice is influenced by the fact that the input ratio is a post-synaptic attribute, in the same spirit as the post-synaptic scaling mechanism found by Turrigiano and coworkers [11, 27]. Let fluctuations of the input ratio about the target input ratio be $\Delta\eta(i;t) = \eta(i;t) - 1$. Let the physical distance between nodes i and j be D(i,j).

We next need equations governing the time evolution of S(i;t) and P(i,j;t). We wish to scale S(i;t) and P(i,j;t) down or up depending on whether each node is firing too frequently or too infrequently, and whether connectivity to that node is too high or too low. Our model for dynamical homeostasis then consists of the equations:

$$\frac{\partial}{\partial t} S(i;t) = -[k_{11}\, \Delta f(i;t) + k_{12}\, \Delta\eta(i;t)]\, S(i;t), \qquad \text{Eq (4)}$$

$$\frac{\partial}{\partial t} P(i,j;t) = -[\, k_{21}\, \Delta f(i;t) + k_{22}\, \Delta\eta(i;t) + k_D\, D(i,j)]\, P(i,j;t). \qquad \text{Eq (5)}$$

The rate constants $k_{11}, k_{12}, k_{21}, k_{22}$ set the timescales for scaling of S(i;t) and P(i,j;t) in response to fluctuations of the relative firing rate and input ratio. For convenience we can define a matrix **K** where the elements of the matrix are the rate constants $k_{ij}$ above. The rate constant $k_D$ controls a distance-dependent cost factor. Larger values of $k_D$ increase the cost of maintaining a connection between two nodes over a period of time, which is greater for nodes that are far apart.

We do not claim that Eqs (4)-(5) represent the only possible algorithm for regulating the spontaneous and stimulated firing probabilities, and we do not expect these equations to hold for extreme conditions, such as when all connections are severed. We adopt these equations only because they are particularly simple and provide a convenient starting point.

Hebbian learning can be incorporated by increasing P(i,j;t) by a factor $1 + C_H$ if firing of node j in the time interval $(t - \delta t, t]$ is followed by firing of node i in the next time interval $(t, t + \delta t]$. We refer to $C_H$ as the Hebbian learning factor. This type of Hebbian learning models long-term potentiation (LTP). Long-term depression (LTD) can be modeled by reducing P(i,j;t) by a factor of $1 - C_H$ if firing of node i in the time interval $(t, t + \delta t]$ is not preceded by firing of node j in the prior time step. Spike-timing dependent plasticity (STDP) can be modeled by combining the criteria for both LTP and LTD.

Equations (4) and (5) can be easily integrated numerically. Because S(i;t) and P(i,j;t) represent probabilities, their values are reset to one if their calculated values ever exceed one. Every node is initially allowed to be connected to every other. The total number of nodes is taken to be N = 64, with 8 rows of 8 nodes in a square lattice with lattice constant *a* (that is, *a* is the distance from one node to its nearest neighbor). The timestep is taken to be $\delta t = 4$ msecs. A refractory period of 20 msecs is imposed after every firing at each node. The target firing interval is $\tau_o = 6.25$ secs. These parameters are chosen to be in agreement with experiment [14, 15].

**Results**

Let us first study the stability properties of the dynamical homeostatic equations, Eq (4)-(5). We have performed extensive computer simulations, allowing the rate constants $k_{ij}$ and the Hebbian learning factor $C_H$ to take on all possible permutations of the values of 0, $2 \times 10^{-5}$ and 0.01, using LTP, LTD and STDP versions of Hebbian learning. There are $3^5$

= 243 possible permutations for each type of Hebbian learning. If convergence is defined numerically to be such that <f(i;t)> and <η(i;t)> both approach unity to within 5% within a total simulation time of 50 million timesteps, then the only states that converge are those listed in Table 1. By visual inspection, it is generally clear which simulations will converge by 3 to 5 million timesteps, but we continue every simulation for at least 50 million timesteps and occasionally up to 300 million timesteps. Note that most choices for $k_{ij}$ and $C_H$ resulted in unstable systems. To summarize Table 1, it appears that convergence requires the following necessary conditions to be met: (1) the spontaneous firing probability must be greater than zero, (2) scaling of the spontaneous firing probability must be dominated by firing rate homeostasis (the $k_{11}$ term), (3) scaling of the connection strengths must be dominated by critical homeostasis (the $k_{22}$ term), (4) "off-diagonal" terms are permissible as long as $k_{11} k_{22} > k_{12} k_{21}$, (5) the timescale for critical homeostasis is at least as fast as Hebbian learning ($k_{22} \geq C_H$), (6) critical homeostasis primarily affects scaling of connection strengths, not firing rate ($k_{22} > k_{12}$), and (7) critical homeostasis occurs on a faster timescale than firing rate homeostasis ($k_{22} >> k_{11}$ converges fastest but sometimes $k_{22} = k_{11}$ may also converge).

We were not able to derive these stability and convergence criteria analytically for the general case. However, if we assume that simultaneous multinodal activations are rare (on average there are 1562 timesteps between each firing event for each node), that this solution does not depend on initial conditions for F(i;t) or A(i;t) (so that the steady state is stable to perturbations), that every node has the same steady state values for firing rate and input ratio (the system is homogeneous), that fluctuations in the zero-time time-correlation between the connection strength and activity are small (i.e., it takes finite time for connection strengths to be adjusted up or down), and that $k_D = 0$ (no distance-dependent connectivity cost factor), then we show in the Appendix that convergence to steady state requires that $S(i;t) > 0$ and $\det(\mathbf{K}) > 0$, where $\det(\mathbf{K}) = k_{11} k_{22} - k_{12} k_{21}$. These two requirements are equivalent to the stability criteria (1)-(4) above. Furthermore, the only stable steady state under these conditions is that for which firing rate and critical homeostasis are achieved, i.e., where $<f(i;\infty)> = <\eta(i;\infty)> = 1$.

The assumptions listed above are all reasonable for our system except for the assumption that $k_D = 0$. Realistic biological systems should have a non-zero distance-dependent connectivity cost factor. However, we will show later in this section that simulation results depend only weakly on this cost factor as long as it is smaller than some particular value. We take up this issue again later in this section; for now, we ignore the term in $k_D$.

It is instructive to look at sample states that did not converge, and to compare them to states that do converge. In Fig 1, we show that firing rate homeostasis by itself cannot guarantee critical homeostasis ($k_{11}$ and $k_{21} > 0$, $k_{12} = k_{22} = 0$), since $\det(\mathbf{K}) = 0$ in this case. This result is not unexpected, since critical homeostasis represents a constraint distinct from firing rate homeostasis, and there is no reason to expect that firing rate homeostasis alone should guarantee critical homeostasis.

In Fig 2, we show that states with $k_{12}$ and $k_{21} > 0$, and $k_{11} = k_{22} = 0$ are also unstable, because $\det(\mathbf{K}) < 0$ in this case. If the initial $S(i;t)$'s are too low, these states tend to fall into global silence with zero firing rates and high input ratios $\eta(i) \approx N$. If the initial $S(i;t)$'s are too high, these states tend to approach tonic hyperactivity with low input ratios, $\eta(i) << 1$. When $\det(\mathbf{K}) < 0$, there appears to be no mechanism for the S(i;t)'s to find an optimal value such that firing rates and input ratios are stable about their target values.

In Fig 3, we show that when $\det(\mathbf{K}) > 0$, stable states are found that maintain relative firing rates and input ratios that fluctuate about unity. Interestingly, not every state with $\det(\mathbf{K}) > 0$ converges to the correct target values. These states appear to depend on initial conditions. It appears from our exploration of numerical examples that additional criteria for convergence to target values are that $k_{22} \geq C_H$, that $k_{22} > k_{12}$, and that $k_{22} \geq k_{11}$ (see Table 1). Although we were unable to prove this analytically, it may be that these additional criteria remove the dependence on initial conditions.

The numerical convergence condition $k_{22} \geq C_H$ states that scaling of the connection strengths must not be much slower than the timescale for Hebbian learning, because otherwise Hebbian learning would dominate critical homeostasis and make critical homeostasis impossible. If the condition $k_{22} \geq C_H$ were not satisfied, then repeated Hebbian learning would eventually destabilize the network, causing either network over- or under-connectivity. Conversely, if $k_{22} \ll C_H$, then the neural network would have stable connectivity but learning would be very slow. Optimal learning with stable connectivity is achieved when $k_{22} \approx C_H$. A caveat, however, is that it is acceptable to allow $C_H$ to be slightly larger than $k_{22}$ for a certain period of time (i.e., while the animal is awake during which speed of learning is advantageous) but then turning $C_H$ down or even off for another period of time (i.e., when the animal sleeps), to allow critical homeostasis to catch up to learning [compare Ref 28].

Why $k_{22}$ must be greater than $k_{11}$ for convergence to occur is not obvious to us. However, an evolutionarily favorable consequence of this criterion is that the dynamics of changes in connectivity is then unconstrained by firing rate homeostasis. That is, if firing rate homeostasis enters on a longer timescale than critical homeostasis, and if the timescale for critical homeostasis and Hebbian learning are similar, then firing rate homeostasis is less likely to interfere with the speed at which an animal can learn.

The stability of our model to fluctuations in the firing rate and branching ratio seems to depend strongly on the behavior of the spontaneous firing probability $S(i;t)$, even though steady-state values of $S(i;t)$ are very small. In our examples, the average $S(i;t)$ is on the order of $1 \times 10^{-5}$ to $8 \times 10^{-5}$. With a timestep of $\delta t = 4$ msecs and target firing interval of $\tau_o = 6.25$ secs, the proportion of total firing due to spontaneous firing then comes out to 2 to 12%. That is, even though most of the activity of our system is due to connectivity-related activity, nonetheless non-zero spontaneous activity is necessary for system stability.

We do not have experimental values for $\mathbf{K}$, $C_H$ or $k_D$. These parameters will depend on the size of the electrodes and the number of neurons each electrode overlies. As discussed earlier, these parameters do not have a simple relationship to analogous neuronal properties. For instance, the nodal spontaneous firing probability will depend on both the local neuronal spontaneous firing probabilities and also on the strengths of local neuronal connectivities, while nodal connection strengths will depend on longer-distance connections between neurons underlying different microelectrodes. Further, the timescales of neuronal dynamics will not translate directly to nodal timescales. Nonetheless, we can make a few statements about their relative magnitudes. First, the rate constant $k_{22}$ should be on the same order of magnitude as the Hebbian learning factor $C_H$, as discussed above, in order that neither dominates the other. Second, all of these timescales must be shorter than the simulation timestep; if this were not true, then we would need to choose a smaller simulation timestep. In effect, this means all the $k_{ij}$'s must be much smaller than one, in units of $1/\delta t$. Third, we expect the rate constant $k_{11}$ to be much slower than $k_{22}$, so that firing rate homeostasis infringes as little as possible on the ability of the neural system to react quickly to the environment. We found above that such a choice also appears to be necessary for the firing rate and input ratio to converge to the correct values.

To demonstrate the effect of varying $k_D$, let $k_{11} = 2 \times 10^{-5}$, $k_{12} = 0$, and $k_{21} = k_{22} = C_H = 0.01$. We integrated Eqs (4)-(5) using $k_D$ equal to 0, $10^{-6}$, $10^{-5}$, $10^{-4}$, $10^{-3}$, and $10^{-2}$, with $k_D$ in units of $1/(a\, \delta t)$ where $a$ is the lattice constant, $\delta t = 4$ msecs is the timestep, and a total of 50 million timesteps are taken per simulation. For the representative case of $k_D = 10^{-5}$, the average input ratio comes out to $\eta = 0.99$ with standard deviation 0.22, and the relative firing rate is 1.01 with standard deviation 0.26.

When $k_D$ is between $10^{-5}$ and $10^{-3}$, the distribution of avalanche sizes $G_A(n)$ as a function of avalanche size n (Fig 4a) is suggestive of a power law with an exponent close to -1.5. At smaller values of $k_D$, the distribution $G_A(n)$ is still suggestive of a power law but there is an upturn at large avalanche sizes with greater than expected frequency of these larger avalanche sizes. Values of $k_D$ larger than $10^{-3}$ destroy the power law dependence, replacing it by a decreasing exponential with a cutoff.

In Fig 4b, we show the relative strength of connectivity as a function of distance. As $k_D$ increases, the connectivity strength drops off very quickly as a function of distance. By $k_D = 10^{-3}$, next-nearest neighbor connectivities are 5 orders of magnitude weaker than nearest neighbor interactions, and yet the corresponding plot in Fig 4a shows that avalanche sizes can still span almost the entire network of 64 nodes. However, by $k_D$ = $10^{-2}$, there are no connections beyond nearest neighbors, and in this case, Fig 4a shows that avalanche sizes are sharply curtailed, to a maximal size of less than 10 nodes. These results suggest that being able to activate large avalanche sizes requires at least some connections beyond nearest neighbors.

The rate constant $k_D$ is related to a lengthscale L = $1/(k_D \tau_o)$. This lengthscale is that at which the increased cost of maintaining a long-distance connection becomes apparent on the timescale of the target firing rate. When $k_D$ is between $10^{-5}$ and $10^{-3}$, L is between 64 and 0.64, respectively, in units of the lattice constant *a*. This range corresponds to the lengthscales of our 8×8 system. Thus the results of Fig 4a and Fig 4b suggest that as long as L $\geq$ *a*, connectivities extend beyond nearest neighbors and activation patterns can spread to cover the entire array.

Fig 4c shows the averaged values of the input ratio, relative firing rate and spontaneous firing probability as a function of $k_D$. As long as $k_D < 10^{-2}$, it is possible to maintain the average input ratio very near unity. That is, critical homeostasis is not very sensitive to the distance-dependent cost factor, until the cost of maintaining connectivity is so high that there are no connections at all beyond nearest neighbors. At larger values of $k_D$, it is no longer possible to maintain critical homeostasis, although firing rate homeostasis is still maintained.

The weak dependence of critical homeostasis on $k_D$ for $k_D < 10^{-2}$ allowed us to ignore $k_D$ in the stability and convergence analysis performed in the Appendix.

Fig 4d shows the distribution of connection strengths as a function of connection strength P. As long as $k_D < 10^{-2}$, the general shape of this distribution is of a tall peak near P = 0 with a long, flat tail or a smooth hump extending out to the maximal strength of P = 1. For the representative plot of $k_D = 10^{-5}$, about 97% of connection strengths are less than P = 0.001. Thus the connectivity pattern, for $k_D < 10^{-2}$, is generally sparse.

Connectivity that is sparse and consists of mostly nearest neighbor connections is less costly for biological neural systems to maintain. When such networks are nonetheless capable of activating large-scale clusters spanning almost the entire network, one is able to reduce metabolic cost (by retaining fewer long distance connections) without giving up much information processing power. This favorable situation is present for our examples with $k_D < 10^{-2}$, and is thus relatively insensitive to $k_D$, as long as the lengthscale requirement, $L \geq a$, is satisfied. Such a situation is suggestive of so-called small world connectivity [18-21]. Small world connectivity has been shown to have precisely this property of being able to maintain global connectivity even with mostly local connections. Furthermore, the ability to maintain global connectivity is relatively insensitive to the percentage of long-distance connections, until this percentage is very nearly zero. This relative insensitivity to the percentage of long-distance connections is also demonstrated by our model qualitatively (Fig 4ab).

Finally, in addition to the specific permutations in the values of the $k_{ij}$'s above, we also randomly sampled other values of $k_{ij}$. We did not find any exceptions to the stability and convergence criteria discussed above. What appeared most important are not the absolute values of the $k_{ij}$'s, but rather their relative values. The absolute values control the absolute timescales of the dynamics, but the relative values are key for stability and convergence.

**Discussion**

We have explored the requirements of a simple biologically plausible neural system model that incorporates (1) spontaneous nodal firing acivity, (2) stimulated nodal firing activity, (3) firing rate homeostasis, (4) critical homeostasis, (5) Hebbian learning, and (6) a distance-dependent connectivity cost function. Our goals were to investigate first whether it is possible to construct such a model, and second whether there are algorithmic consequences of imposing both firing rate and critical homeostasis for this particular model. Our primary findings are that we were indeed able to construct such a model, and we found that not every possible set of parameters for our model is stable.

In particular, within the limits of our simple model, we found from extensive computer simulations that stability and convergence of the system to target firing rates and connectivity require that (1) the spontaneous firing probability must be greater than zero, (2) scaling of the spontaneous firing probability must be dominated by firing rate homeostasis (the $k_{11}$ term), (3) scaling of the connection strengths must be dominated by critical homeostasis (the $k_{22}$ term), (4) "off-diagonal" terms are permissible as long as $k_{11} k_{22} > k_{12} k_{21}$, (5) the timescale for critical homeostasis is at least as fast as Hebbian learning ($k_{22} \geq C_H$), (6) critical homeostasis primarily affects scaling of connection strengths, not firing rate ($k_{22} > k_{12}$), and (7) critical homeostasis occurs on a faster timescale than firing rate homeostasis ($k_{22} \geq k_{11}$). The first four stability conditions can also be derived analytically, under certain circumstances discussed as above and in the Appendix.

A secondary finding of our simulations is that by making only the simple, biologically plausible assumptions above, the resulting model neural system is able to activate cluster or avalanche sizes ranging from the smallest to the largest possible sizes (the system is scale-free), and this property is maintained even when most connection strengths are very weak and most are very short distance (i.e., when $k_D > 0$). This finding suggests that neural systems that satisfy our basic assumptions have available to them near-maximal computational power at low metabolic cost, over a fairly wide range of parameters.

We do not know if our conclusions are generalizable to other specific neural system models, nor whether real biological systems have the same stability and convergence requirements. Nonetheless, our results suggest that real biological systems may have similar constraints which must be satisfied in order for these systems to function optimally. If such constraints exist, then they will be important to discover and characterize, because real life failure to satisfy these constraints will represent disease states. For instance, neural tissue that is persistently underconnected will have difficulty activating large-scale circuits, and may then result in learning disorders and mental retardation. Understanding the nature of these constraints may also help us to manipulate them in times of physiological stress, to protect patients after a brain insult.

As an example, we simulated a seizure on our system by forcing 12 out of 64 nodes to fire at rates far above the target rates for a period of time (Fig 5). From simulations, we found that, during a seizure, hyperactive firing at one set of nodes drives the spontaneous firing probability of all system nodes to lower values (Fig 5a). The connectivities are also driven to lower values. At the end of the seizure, both the spontaneous firing probabilities and the connectivities begin to recover, but the connectivities recover faster because $k_{22} > k_{11}$. If it happens that $k_{22} \gg k_{11}$, then the connectivity will actually overshoot for a period of time, and the post-seizural state will be supercritical even though activity is depressed (Fig 5bc). In our example, the post-seizural supercritical state lasts on the order of 50 million timesteps (about 55 hours). We conjecture that this post-seizural supercritical state plays a role in epileptogenesis, by increasing the chance of creating a large-scale recurrent activation pathway. That is, an over-connected state increases the chance that the neural system may learn to activate a pathway that comes back on itself, and not only that, but that this pathway may encompass a large number of nodes in the system. The recurrent nature of such a pathway gives rise to persistent self-activation, while the involvement of a large number of nodes gives rise to large-scale synchronization, both properties of which are necessary for seizural states. If the post-seizural state is indeed supercritical and epileptogenic, then we suggest that post-seizural interventions that either boost spontaneous firing or that inhibit connectivity-dependent firing will decrease the tendency towards supercriticality, and thus be protective against epileptogenesis. That is, counterintuitively, spontaneous activity and connectivity-dependent activity have opposite effects on epileptogenesis, and thus it is important to distinguish between these two types of activity.

Another example of a disease state is when an area of cortex is suddenly cut off from its neighbors, or "deafferentated" [29]. Deafferentation is a model for traumatic brain injury. Immediately after deafferentation, neuronal activity drops precipitously. Epileptiform discharges then later appear. Why does this happen? Assuming that the target firing interval $\tau_o$ is the same in intact brain as in deafferentated brain, our model predicts that the baseline spontaneous firing probability needed to maintain the target firing rate should be smaller for intact brain than for deafferentated brain (Fig 6a). The reason is that spontaneous activity can propagate to other nodes, and hence, if there are more nodes, lower spontaneous activity is needed to maintain the same level of activity. With acute deafferentation the firing rate therefore immediately drops (Fig 6b). In response to the drop in firing rate, connectivities increase to supercritical levels and remain supercritical until the spontaneous firing probability reaches its new steady state value (Fig 6b). In our example, the post-deafferentation supercritical state lasts for 9 hours. Again, we hypothesize that this period of supercriticality is one during which epileptogenesis may occur, and that interventions that either boost spontaneous firing or that inhibit connectivity-dependent firing will decrease the tendency towards supercriticality, and thus be protective against epileptogenesis. In addition, if for some reason there is a ceiling on maximal spontaneous firing probabilities, such that the deafferentated nodes are blocked from reaching these critical spontaneous firing rates, then it is possible to produce an indefinitely persistent supercritical state.

Latham et al. [30] have also studied the role of spontaneous activity in neural networks. This model is somewhat different from ours, involving a heterogeneous population of cells which includes a fraction of endogenously active cells. They find that increasing the fraction of endogenously active cells produces firing patterns that are more regular and of lower frequencies. Systems with too few endogenously active cells are either silent or fire in bursts of high frequency. Our system is homogeneous and describes not individual cells but groups of cells, each group represented by a node. In our model, when the spontaneous firing probabilities $S(i;t)$ are too low for a prolonged period, for whatever reason, the firing rates also become too low. In partial compensation, the connection strengths $P(i,j;t)$ then rise to supercritical levels. What tends to happen in this situation is that the baseline firing rate is still too low, but when there is spontaneous activity, that activity quickly spreads to many nodes, in a burst, because the connectivity is supercritical. During the period of the activity burst, the firing frequency would appear very high. If the $S(i;t)$'s are allowed to recover (that is, as they slowly increase), then the $P(i,j;t)$'s are scaled back down. When the connectivity returns to critical or subcritical levels, then spontaneous firing of a node is less likely to spread quickly to a large number of other nodes. The firing pattern then becomes more regular. If we take our spontaneous firing probability $S(i;t)$ to be analogous to the fraction of endogeneously active cells of Latham et al., then our conclusions are concordant with theirs.

Regarding spontaneous firing, we also found that spontaneous activity accounts for only 2 to 12% of total activity, and yet stability about target firing rate and connectivity requires that spontaneous activity must be greater than zero. Perhaps this requirement for stability is the reason why the brains of higher animals (i.e., those capable of learning and adapting to the environment) are never entirely quiet, even at rest and even in sleep.

Two of us have previously attempted to construct a model exhibiting both firing rate homeostasis and critical connectivity, but assuming only a mechanism for firing rate homeostasis [31]. This earlier model is the same as the model presented here but with no homeostatic mechanism for connectivity (i.e., $k_{21}$ and $k_{22}$ were both equal to zero). In addition, the earlier model was not capable of Hebbian learning ($C_H = 0$). It was found that as long as the rate constant for scaling of the spontaneous firing probability is smaller than that for the connection strength ($k_{11} < k_{21}$), then critical connectivity was approximately maintained, with best results for $k_{11}/k_{21} = 0.5$. However, when Hebbian learning was turned back on ($C_H > 0$), no stable state that converged on target firing rates and critical connectivity could be found despite extensive exploration of parameter space (Hsu and Beggs, unpublished). It was the failure to find such a state, when Hebbian learning was turned on, that led us to consider whether homeostasis of connectivity may be a homeostatic principle in its own right, distinct from that of firing rate homeostasis.

Abbott and Rohrkemper [32] have proposed an elegant "growth" model that essentially allowed connectivity to grow or shrink so as to achieve firing rate homeostasis. Power-law behaviors were then seen suggestive of critical connectivity. That is, critical behavior was seen in this model assuming only firing rate homeostasis, without having to invoke critical homeostasis explicitly. As in our own earlier effort, however, this model does not incorporate Hebbian learning. It would be interesting to know whether this growth model remains capable of critical behavior after Hebbian learning is turned on.

Sullivan and de Sa [33] proposed a simple model capable of both Hebbian learning and firing rate homeostasis. However, they did not investigate whether their model exhibits critical connectivity, and their model is not meant to be spontaneously active. The model of Yeung et al. [34] incorporating a calcium-dependent plasticity is appealing for its underlying realism, but it similarly does not investigate issues of critical connectivity. Others have investigated mechanisms of firing rate homeostasis, usually in the form of constraints [2, 3]. A modification of spike timing dependent plasticity learning rules can also achieve activity homeostasis [35]. However, this model does not explicitly incorporate critical homeostasis, and it is not clear that critical homeostasis would arise as a natural consequence of the dynamics. There are also models which are capable of self-organized criticality based purely on local learning rules [36]. Such models are attractive in that they do not need to suppose nonlocal learning rules. In contrast, we assume that each node must sense the total input to that node, which is a nodal-wide property and not a strictly local (i.e., synaptic) property. However, the experimental finding of multiplicative synaptic scaling [11] and of active and passive backpropagation of action potentials into the dendritic tree [37-39] show that nonlocal neuron-wide mechanisms do exist, and suggest that nodal-wide mechanisms may also exist. We leave it to future experiments to determine whether critical homeostasis really involves only purely synaptic-level learning rules.

The precise magnitudes of the rate constants $k_{ij}$ are not known, as experiments have not been framed in terms of determining these rate constants. Of particular interest are not simply the absolute magnitudes of these rate constants, but their relative magnitudes, because stability and convergence depends on the relative magnitudes and not so much on the absolute magnitudes. We suggest that it may be interesting to design experiments to look at these timescales, to see if these timescales are constrained in the ways we have described above, and also to see if a simple scaling algorithm is found as in Eqs (4) and (5).

In terms of orders of magnitudes of timescales, the synaptic scaling mechanism of Wierenga, et al. [11] is a slow process, with a timescale of hours to days. Since critical homeostasis must occur on the same timescale as Hebbian learning, in order that neither critical homeostasis nor Hebbian learning dominates the other, we must look elsewhere for a fast biomolecular mechanism for critical homeostasis. In addition to being fast, such a mechanism must also be nonlocal, i.e., not restricted to the level of individual synapses, because the input and branching ratios are nonlocal properties requiring simultaneous knowledge of total input and output weights across an entire node. Here we mention three examples of candidate fast, nonlocal mechanisms: (a) When homosynaptic LTP (or LTD) is induced in the intercalated neurons of the amygdala, compensatory heterosynaptic depression (or facilitation) is observed such that the total synaptic weight of a given neuron remains constant [40]. The counterbalancing heterosynaptic response is suggestive of the scaling mechanism of Eqs (4) and (5). This mechanism depends on the release of intracellular calcium stores and has a timescale of minutes. (b) The phenomenon of backpropagation of action potentials into the dendritic tree, as seen in certain neocortical and hippocampal neurons [37-39], allows widely distributed numbers of synapses to receive nearly simultaneous information about neuronal output. This information is conjectured to play a role in LTP, LTD and STDP, but might conceivably also be used for critical homeostasis. (c) A third mechanism may involve the interaction of principal output neurons with local interneurons. It may be that a certain subset of local interneurons can sense both the total input into and total output out of a local community of output neurons. This information might then be used to modulate either the input or branching ratio of that group of output neurons. In support of this possibility, blocking interneurons with bicuculline produces a dramatic increase in the branching ratio within minutes (Beggs unpublished).

Which of these three mechanisms, if any, are related to critical homeostasis remains to be seen. Nonetheless, that these mechanisms are all fast and nonlocal suggests that an appropriate mechanism for critical homeostasis is at least biologically plausible.

We reiterate that there are theoretical reasons why critical homeostasis should exist in systems that maintain capacity for continual learning, because neural systems that operate far from criticality are inefficient and unreliable. Thus, we hope to encourage future experiments to determine which neural systems are capable of critical homeostasis, how tightly critical homeostasis is maintained, and what mechanisms underlie it.

Desirable improvements in our model include generalization to include higher order multinodal interactions (e.g., terms of the form $P(i,j,k;t)$ as discussed in the Methods section), and inclusion of non-Markovian memory effects in the pairwise connection strengths $P(i,j;t)$. That is, $P(i,j;t)$ represents the conditional probability that firing of node j at time t causes firing of node i in the next time instant, but in our model there is no further effect at timesteps beyond the adjacent timestep. We refer to this type of connectivity as being Markovian.

Non-Markovian connectivity would allow a node to remember from which other nodes it received input during timesteps beyond the immediately preceding timestep. Such memory would allow the formation of more complex spatiotemporal patterns of activity. We have not extensively investigated the effects of non-Markovian memory, but we have experimented with introducing some non-Markovian effects with a memory kernel, replacing Eqs (1) and (2) with the following:

$$A(i;t + \delta t) = 1 - [1 - S(i;t)] \prod_{j=1}^{N} [1 - G(i,j;t)]. \qquad \text{Eq (6)}$$

$$G(i,j;t) = \sum_{t'=0}^{\infty} G_0(t') P(i,j;t-t') F(j;t-t') \qquad \text{Eq (7)}$$

Here $G(i,j;t)$ represents the conditional probability that node i fires at time t due to activity from node j at any prior time, summed over all prior times t' with a non-Markovian weighting factor $G_0(t')$. We choose $G_0(t')$ to have the form of a delta function at short times plus a Gaussian centered at longer times (see Fig 7). The delta function at short times represents Markovian connectivity, while the Gaussian at longer times represents one particular form of non-Markovian connectivity. The branching ratio and input functions are then redefined to include non-Markovian effects as:

$$\eta(i;t) = \left[ \sum_{j=1}^{N} P(i,j;t) \right] \left[ \sum_{t'=0}^{\infty} G_0(t') \right]. \qquad \text{Eq (8)}$$

$$\sigma(i;t) = \left[ \sum_{j=1}^{N} P(j,i;t) \right] \left[ \sum_{t'=0}^{\infty} G_0(t') \right]. \qquad \text{Eq (9)}$$

Using Eqs (6)-(9), we find through computer simulations that the same stability criteria as described above appear to be preserved.

The significance of Markovian vs non-Markovian connectivity is that Markovian systems are not as good for the formation of sequential memory, wherein specific temporal sequences of spatial patterns are to be learned. To see this, consider what happens if exactly the same stimulus is fed into a Markovian neural system, at intervals longer than the nodal refractory period. In this case, every stimulus appears exactly the same to the neural system, and the neural system has no way to know if it has just seen the stimulus for the first time, the second time, or the $n^{th}$ time. There is no sense of "history" in a Markovian system, no possibility of "historical context". Such a system will respond to the same stimulus in the same way, every time that stimulus is presented, regardless of what may have happened in the meantime. In contrast, a system with non-Markovian connectivity will see each succeeding stimulus a little differently, as modified by recent history. Since we know that some animals are capable of historical context and sequential memory, we predict that these animals, including ourselves, have non-Markovian connectivity.

**Grants:** D.H. thanks the American Epilepsy Society and the National Institutes of Health K12 Roadmap Project number 8K12RR023268-03 for support. J.M.B. thanks the National Science Foundation for support.

**Appendix**

We were not able to find a general closed form expression for a steady state solution of Eqs (4)-(5). However, if we make certain assumptions, then a steady state solution can be found. The assumptions are suggestive, but we do not know if they are necessary conditions for a steady state solution to exist.

The assumptions we make are (a) that simultaneous multinodal activations are rare, (b) that these solutions are independent of the initial condition, (c) that the steady state average firing rate and input ratio are the same for every node, and (d) that fluctuations in the zero-time time-correlation between the connection strength and activity are small. Assumption (a) is satisfied when the system is not persistently supercritical and when $\delta t \ll \tau_o$ where $\delta t$ is the timestep and $\tau_o$ is the target time interval between firings. Assumption (b) states that the steady states solutions are stable to perturbations. Assumption (c) states that the system is homogeneous. Assumption (d) states that it takes finite time for connection strengths to be adjusted up or down. We will ignore the distance-dependent cost factor, $k_D$. This factor is discussed separately in the text. To begin, the dynamical homeostatic equations can be written:

$$\frac{\partial}{\partial t} S(i;t) = -[k_{11} \Delta f(i;t) + k_{12} \Delta \eta(i;t)] \, S(i;t), \qquad \text{Eq (A1)}$$

$$\frac{\partial}{\partial t} \eta(i;t) = -[k_{21} \Delta f(i;t) + k_{22} \Delta \eta(i;t)] \, \eta(i;t), \qquad \text{Eq (A2)}$$

where $\Delta f(i;t) = f(i;t) - 1$ and $\Delta \eta(i;t) = \eta(i;t) - 1$. The latter equation is obtained by summing both sides of Eq (5) over all j. For later convenience, let us define a matrix **K**:

$$\mathbf{K} = \begin{bmatrix} k_{11} & k_{12} \\ k_{21} & k_{22} \end{bmatrix} \qquad \text{Eq (A3)}$$

Next, define the steady state values of $\langle f(i;\infty)\rangle = f_o$, $\langle \eta(i;\infty)\rangle = \eta_o$, and $\langle S(i;\infty)\rangle = S_o$, where $\langle ... \rangle$ denotes a time average over a time much longer than $\tau_o$. These steady state values must be the same for each node because the system is homogeneous. For later convenience, define $\delta f(i;t) = f(i;t) - f_o$, $\delta \eta(i;t) = \eta(i;t) - \eta_o$, and $\delta S(i;t) = S(i;t) - S_o$. Recall that:

$$A(i;t + \delta t) = 1 - [1 - S(i;t)] \prod_{j=1}^{N} [1 - P(i,j;t) F(j;t)]. \qquad \text{Eq (A4)}$$

When simultaneous multinodal activations are rare, as we have assumed in assumption (a), then $A(i;t + \delta t)$ can be expanded to lowest orders in $S(i;t)$ and $F(j;t)$:

$$A(i;t + \delta t) \approx S(i;t) + \sum_{j=1}^{N} P(i,j;t) F(j;t). \qquad \text{Eq (A5)}$$

Note that Eq (A5) represents a recursion relationship which generates a trajectory for the vector $\mathbf{A}(t)$. The firing function $F(i;t)$ is essentially a digitized representation of $A(i;t)$. The parameters that control the trajectory for $\mathbf{A}(t)$ are the vector $\mathbf{S}(t)$, the matrix $\mathbf{P}(t)$, and the initial values for $F(i;0)$. Since we are only interested in stable steady state solutions that do not depend on initial conditions, we will regard the trajectory of $\mathbf{A}(t)$ as being dependent only on $\mathbf{S}(t)$ and $\mathbf{P}(t)$. The variation of $A(i,t + \delta t)$, as $S(i;t)$ and $P(i,j;t)$ are varied, is then given by:

$$\delta A(i;t + \delta t) \approx \delta S(i;t) + \sum_{j=1}^{N} \delta P(i,j;t) F(j;t). \qquad \text{Eq (A6)}$$

Note that $\langle F(i;t)\rangle = \langle A(i;t)\rangle$, and that $f(i;t) = \langle F(i;t)\rangle_a \tau_o/\delta t$, where $\langle ... \rangle_a$ denotes a time average over a time equal to $\tau_o$. Applying assumption (d) above, we have

$$\langle \delta P(i,j;t) F(j;t)\rangle \approx \langle \delta P(i,j;t)\rangle \langle F(j;t)\rangle. \qquad \text{Eq (A7)}$$

Combining Eqs (A6) and (A7) we then have:

$$<\delta f(i;t)> \approx (\tau_o/\delta t) <\delta S(i;t)> + f_o <\delta \eta(i;t)>. \qquad \text{Eq (A8)}$$

Taking the time derivative of both sides in Eq (A8), combining that equation with Eqs (A1) and (A2), and expanding to lowest order in $<\delta S(i;t)>/S_o$ and $<\delta\eta(i;t)>/\eta_o$, we have:

$$\frac{\partial}{\partial t} <\delta f(i;t)> = B_{11} (1 - f_o) + B_{12} (1 - \eta_o) - B_{11} <\delta f(i;t)> - B_{12} <\delta\eta(i;t)>,$$

$$\text{Eq (A10)}$$

$$\frac{\partial}{\partial t} <\delta\eta(i;t)> = B_{21} (1 - f_o) + B_{22} (1 - \eta_o) - B_{21} <\delta f(i;t)> - B_{22} <\delta\eta(i;t)>,$$

$$\text{Eq (A11)}$$

where the matrix **B** is defined as:

$$B_{11} = k_{11} S_o (\tau_o/\delta t) + k_{21} f_o \eta_o,$$

$$B_{12} = k_{12} S_o (\tau_o/\delta t) + k_{22} f_o \eta_o,$$

$$B_{21} = k_{21} \eta_o,$$

$$B_{22} = k_{22} \eta_o. \qquad \text{Eq (A12)}$$

A little manipulation then shows that

$$\begin{bmatrix} <\delta f(i;t)> \\ <\delta\eta(i;t)> \end{bmatrix} = \begin{bmatrix} 1-f_o \\ 1-\eta_o \end{bmatrix} + \exp(-\mathbf{B}\, t) \begin{bmatrix} \delta f(i;0)+f_o - 1 \\ \delta\eta(i;0)+\eta_o - 1 \end{bmatrix} \qquad \text{Eq (A13)}$$

The steady state corresponds to $<\delta f(i;t)> \approx 0$ and $<\delta\eta(i;t)> \approx 0$. Eq (A13) thus shows that a stable steady state exists only if both eigenvalues of **B** are positive and if $f_o = \eta_o = 1$. It is a simple matter to show that the eigenvalues of **B** are $\lambda = b_o \pm \sqrt{b_o^2 - d}$ where $b_o = (B_{11} + B_{22})/2$ and $d = \det(\mathbf{B}) = S_o (\tau_o/\delta t) \eta_o \det(\mathbf{K})$. Therefore, under assumptions (a)-(d) above, the existence of steady state solutions requires $f_o = \eta_o = 1$, $S_o > 0$, and $k_{11} k_{22} > k_{12} k_{21}$.

**Figure legends**

Fig 1. Firing rate homeostasis by itself ($k_{11}$ and $k_{21} > 0$, $k_{12} = k_{22} = 0$) cannot guarantee critical homeostasis. With $k_{11} = 2\times 10^{-5}$ and $k_{21} = 0.01$, the average relative firing rate converges to $f = 1$ but the average input ratio converges to $\eta = 9.8$. With $k_{11} = 0.01$ and $k_{21} = 2\times 10^{-5}$, $f$ converges to one but $\eta$ converges to 57.2. With $k_{11} = 0.01$ and $k_{21} = 0.01$, $f$ converges to one but $\eta$ converges to $6.5\times 10^{-4}$. All values are averaged over the prior one hour of data and over all $N = 64$ nodes. The Hebbian learning factor in these calculations is $C_H = 0.01$. Each timestep is 4 msecs. Note the semi-logarithm scale. A total of 50 million timesteps are taken for each simulation but only the first 20 million timesteps are shown.

Fig 2. Firing rate homeostasis scaling of connection strengths and critical homeostasis scaling of spontaneous firing probabilities result in an unstable system. In these two examples, $k_{11} = k_{22} = 0$, $k_{12} = 2\times 10^{-5}$, $k_{21} = 0.01$, and $C_H = 0.01$. Timestep is 4 msecs. Note the log-log scales. In example (a), the initial $S(i;0)$'s are too small to maintain firing rates and input ratios at unity. When the relative firing rates eventually drops below unity, the connection strengths $P(i,j;t)$ are scaled up and the input ratios rise. As the input ratios rise, the spontaneous firing probabilities are scaled down further, which causes the relative firing rates to drop even more, perpetuating a cycle such that the system eventually becomes silent ($f \approx 0$) despite maximal input ratios ($\eta(i) \approx N$ where $N$ is the total number of nodes). Varying the relative magnitudes of $k_{12}$ and $k_{21}$ does not improve stability. In example (b), the initial $S(i;0)$'s are too large, and the opposite situation

arises, where a cycle is entered such that the spontaneous firing probabilities rise as the input ratios drop. This system then enters tonic hyperactivity with minimal connectivity.

Fig 3. The convergence criterion $\det(\mathbf{K}) > 0$ does not guarantee convergence to the target asymptotic firing rate and input ratio. With $k_{11} = 2\times10^{-5}$, $k_{12} = 0$, and $k_{21} = k_{22} = C_H = 0.01$, both f and $\eta$ converge to unity. However, with $k_{11} = 0.01$, $k_{12} = 0$, and $k_{21} = k_{22} = C_H = 2\times10^{-5}$, f converges to 1.16 while $\eta$ converges to 0.84, even though $\det(\mathbf{K})$ are identical in the two cases. All values are averaged over the prior one hour and over all N = 64 nodes. Timestep is 4 msecs.

Fig 4: Dependence on $k_D$. Example with $k_{11} = 2\times10^{-5}$, $k_{12} = 0$, $k_{21} = k_{22} = C_H = 0.01$, timestep of 4 msecs and total of 50 million timesteps. (a) Log-log plot of the distribution of avalanche sizes $G_A(n)$ vs avalanche size n. Solid line is a power law with exponent of -1.5 shown for reference. (b) Semi-log plot of relative connectivity vs distance, calculated by adding all P(i,j;t)'s at each distance and normalizing area under the curve to unity. Distances are in units of the lattice constant *a*, rounded to nearest integral value. (c) Log-log plot of averaged spontaneous firing probability, input ratio and relative firing rate as a function of $k_D$. (d) Semi-log plot of distribution of connection strengths P(i,j;t) vs connection strength, with total area under the curve normalized to unity.

Fig 5. Simulated seizure for different values of $k_{11}$: a simulated seizure occurs between timesteps $1.25\times10^8$ to $1.5\times10^8$ during which the activation levels A(i;t) of 12 nodes were set to 0.8. In all examples, $k_{12} = 0$, $k_{21} = k_{22} = C_H = 0.01$, and timestep is 4 msecs. (a) The spontaneous firing probability averaged over all nodes, S, declines with seizure onset and slowly recovers after the seizure ends. The rate of recovery is slower for smaller values of $k_{11}$. (b) The relative firing rate as a function of number of timesteps: for $k_{11} = 10^{-6}$ and $k_{11} = 10^{-7}$, the relative firing rate is transiently depressed after the seizure. Note the semilogarithmic scale. (c) The average input ratio $\eta$ as a function of number of timesteps: for $k_{11} = 10^{-6}$ and $k_{11} = 10^{-7}$, the average input ratio becomes supercritical after the seizure ends and remains supercritical for about 50 million timesteps (55 hours).

Fig 6. Simulated acute deafferentation: a system of 100 nodes is suddenly reduced to 10 nodes at timestep $1\times10^8$. Parameters are $k_{11} = 10^{-6}$, $k_{12} = 0$, $k_{21} = k_{22} = C_H = 0.01$, and timestep is 4 msecs. (a) The spontaneous firing probability averaged over all nodes, S, as a function of timesteps. (b) The input ratio $\eta$ and relative firing rate f averaged over all nodes as a function of timestep. Acute deafferentation in this example results in depressed firing rate with supercritical input ratio for 8 million timesteps (9 hours).

Fig 7. Non-Markovian weighting function $G_0(t)$ as a function of time. In this example, $G_0(t)$ consists of a step function at short times plus a Gaussian centered at longer times. The areas under the step function and under the Gaussian are approximately equal.

**Table 1: Numerical tests of stability**

| $k_{11}$ | $k_{12}$ | $k_{21}$ | $k_{22}$ | $C_H = 0$ | $C_H = 2\times 10^{-5}$ | $C_H^* = 0.01$ | $C_H^{**} = 0.01$ |
|---|---|---|---|---|---|---|---|
| $2\times 10^{-5}$ | 0 | 0 | 0.01 | y | y | y | |
| $2\times 10^{-5}$ | 0 | $2\times 10^{-5}$ | $2\times 10^{-5}$ | y | y | | |
| $2\times 10^{-5}$ | 0 | $2\times 10^{-5}$ | 0.01 | y | y | y | y |
| $2\times 10^{-5}$ | 0 | 0.01 | 0.01 | y | y | y | y |
| $2\times 10^{-5}$ | $2\times 10^{-5}$ | 0 | 0.01 | y | y | y | y |
| $2\times 10^{-5}$ | $2\times 10^{-5}$ | $2\times 10^{-5}$ | 0.01 | y | y | y | y |
| $2\times 10^{-5}$ | 0.01 | 0 | 0.01 | y | y | | |

**Table legend**

Table 1. Numerical tests of stability. List of all choices of $k_{ij}$ such that the relative firing rates and input ratios converged to unity. We allowed each $k_{ij}$ and the Hebbian learning factor $C_H$ to take on the values of 0, $2\times 10^{-5}$ and 0.01 in all possible permutations. A timestep of $\delta t = 4$ msecs was taken, for a total of 50 million timesteps for each set of parameters. Other details are as decribed in the Methods section. The states that converged are denoted with a "y". All data refer to LTP results except the column marked $C_H^*$, which denotes identical results for both LTP and LTD, and the column marked $C_H^{**}$, which denotes use of an STDP algorithm.

Fig 1

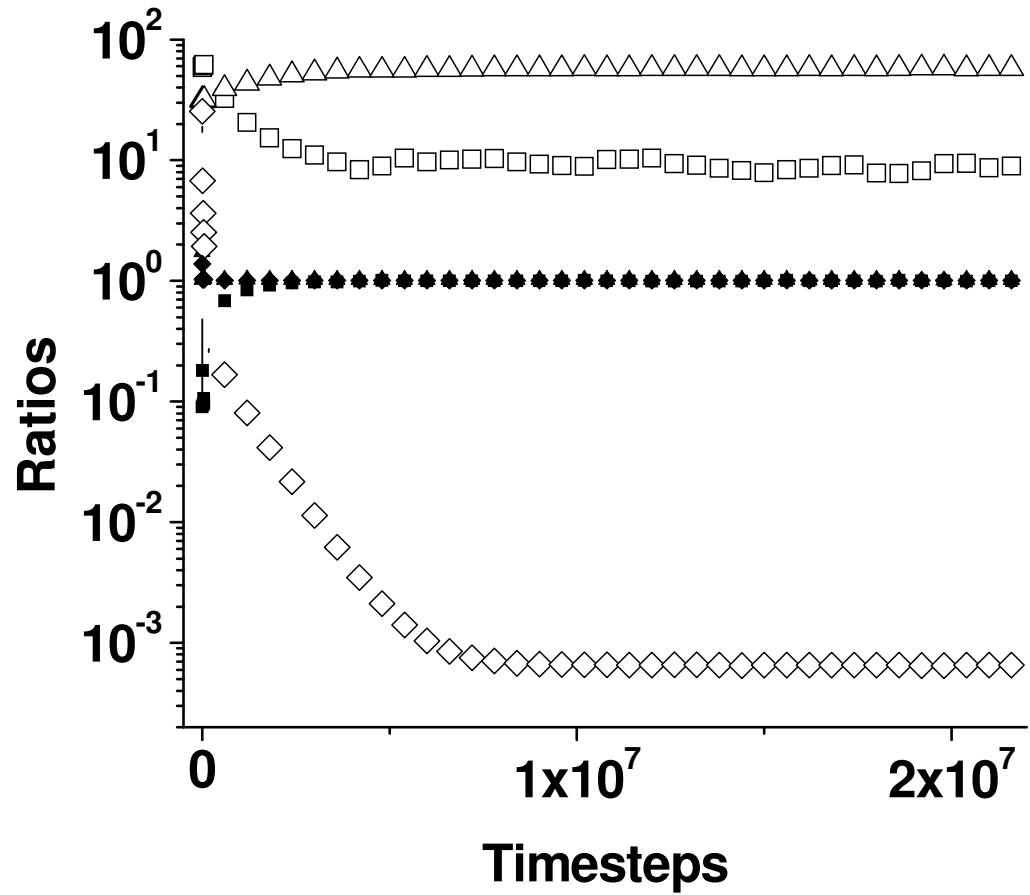

Fig 2a

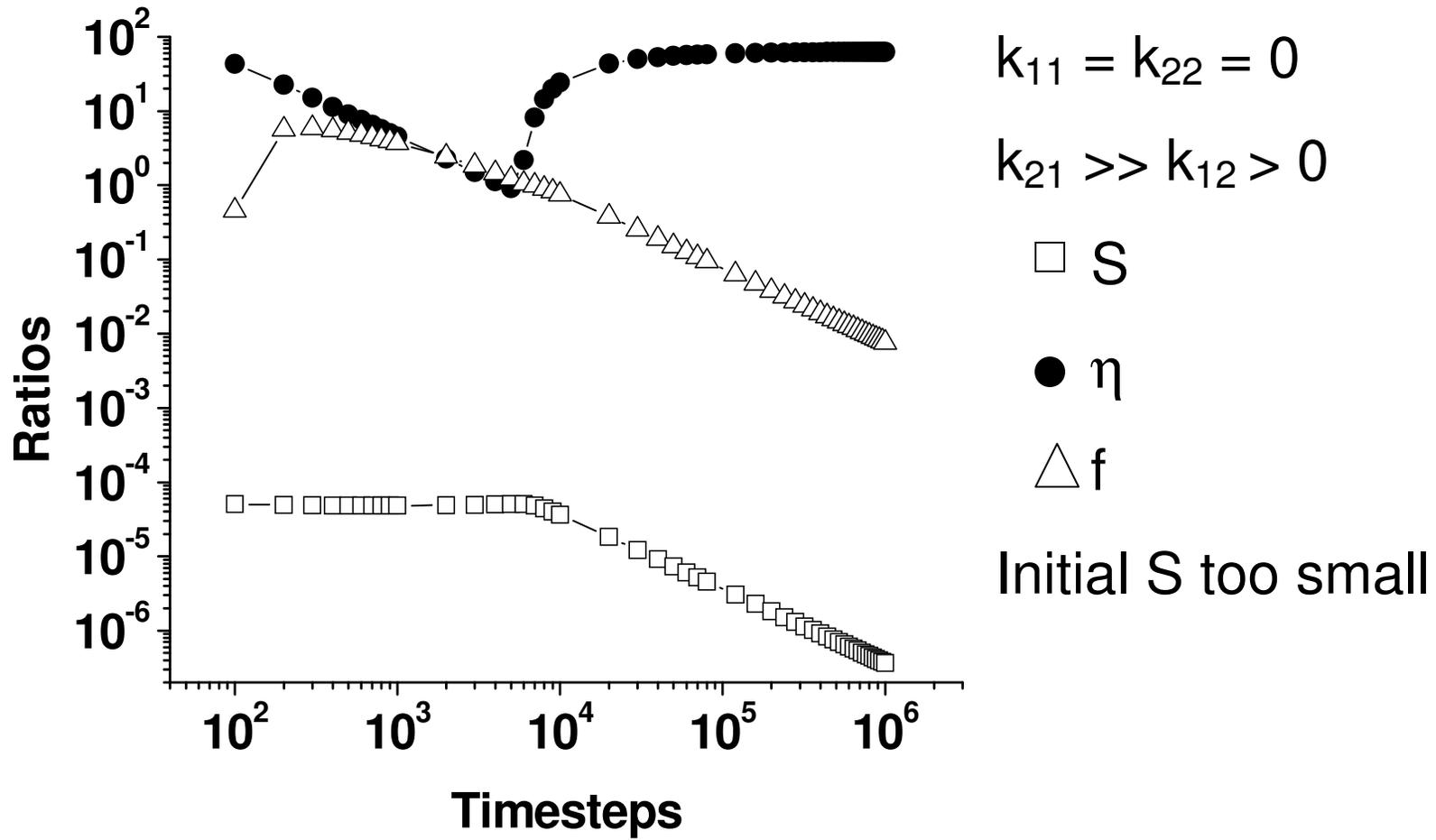

$k_{11} = k_{22} = 0$

$k_{21} \gg k_{12} > 0$

□ S

● η

△ f

Initial S too small

Fig 2b

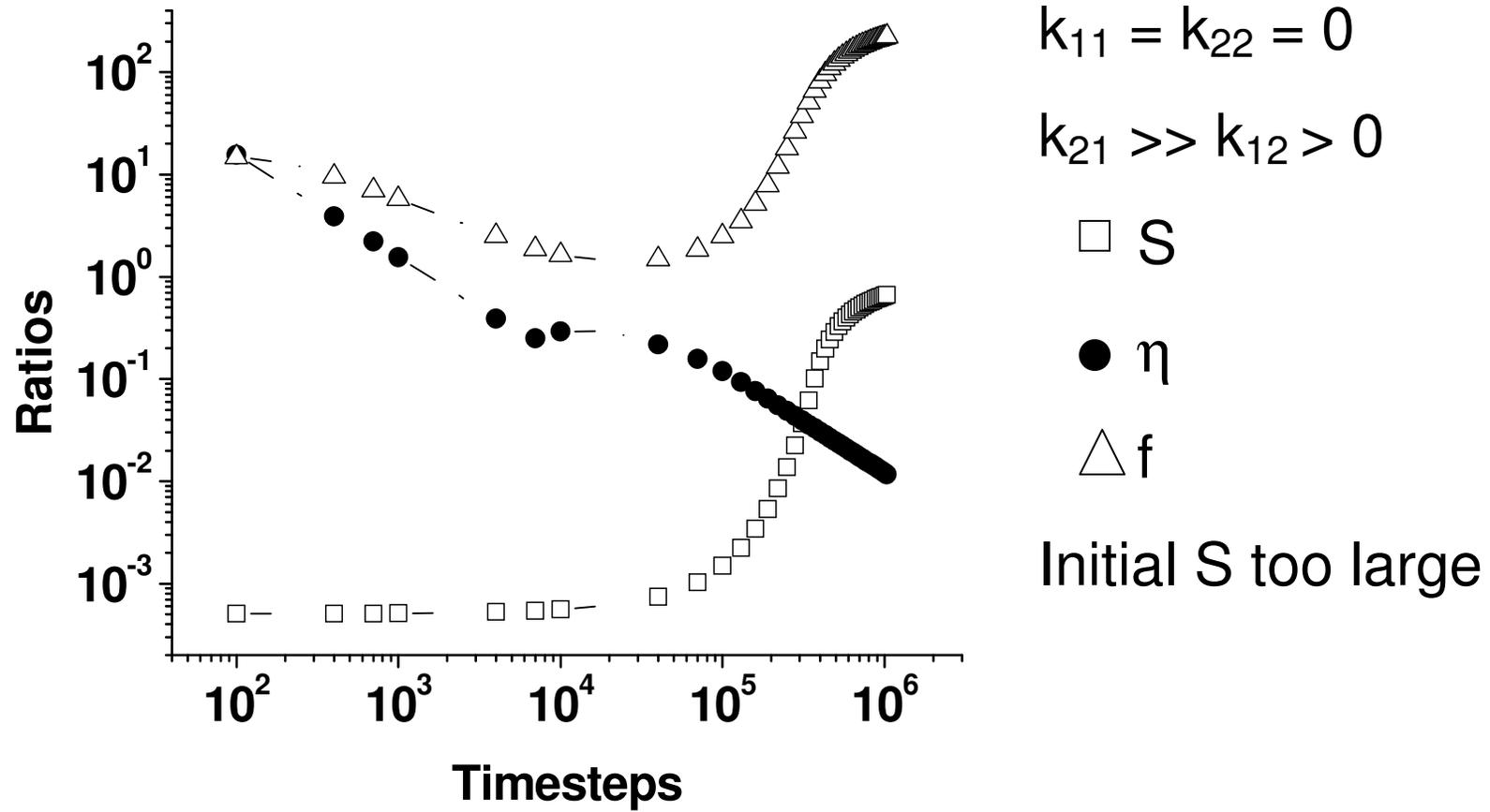

$k_{11} = k_{22} = 0$

$k_{21} \gg k_{12} > 0$

□ S

● η

△ f

Initial S too large

Fig 3

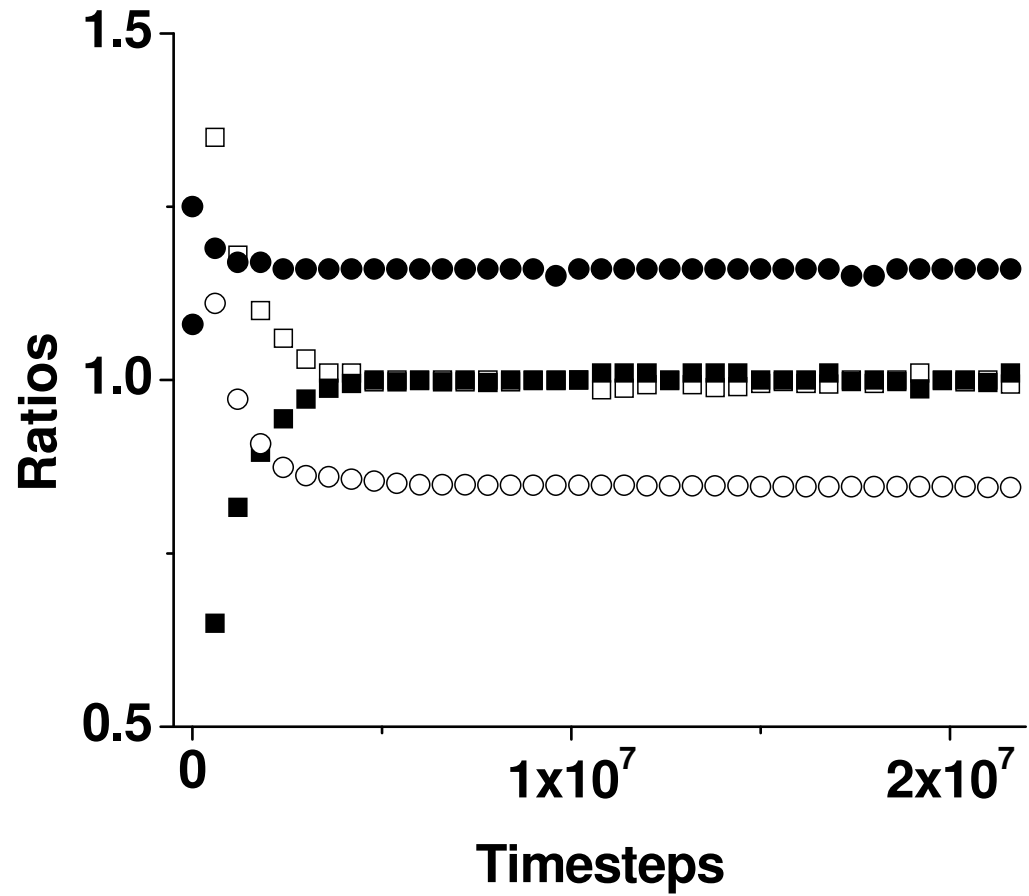

Fig 4a

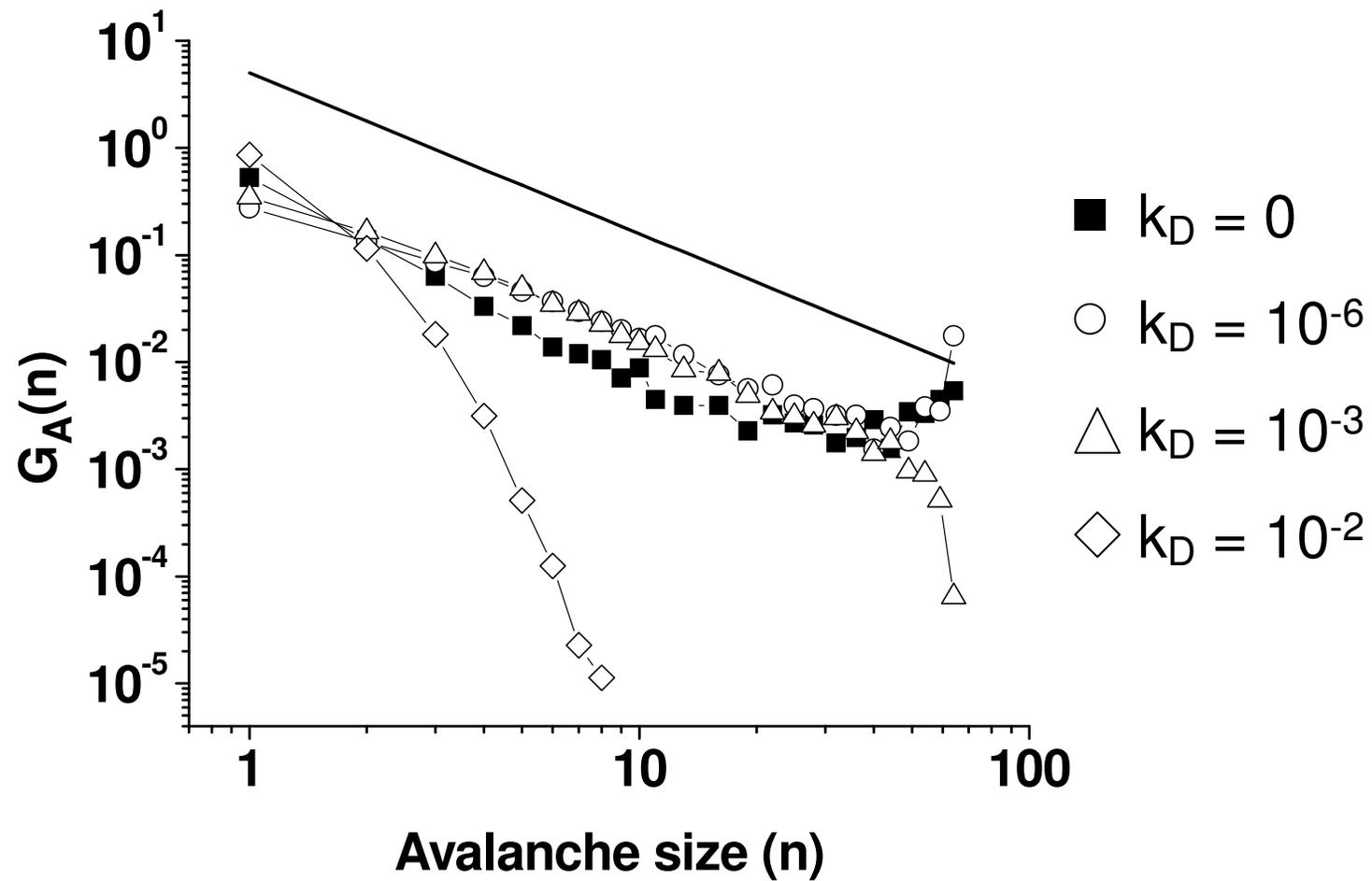

Fig 4b

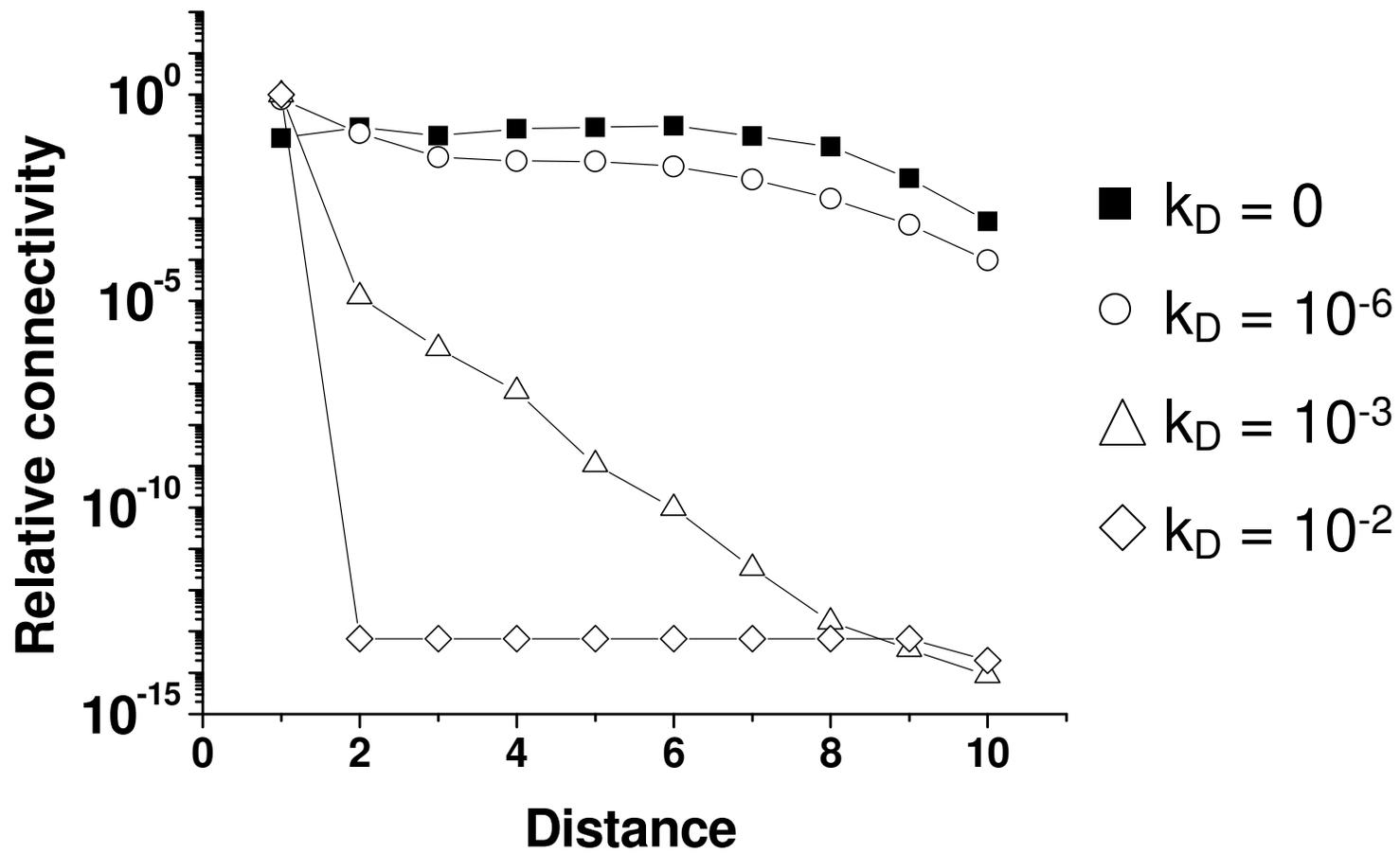

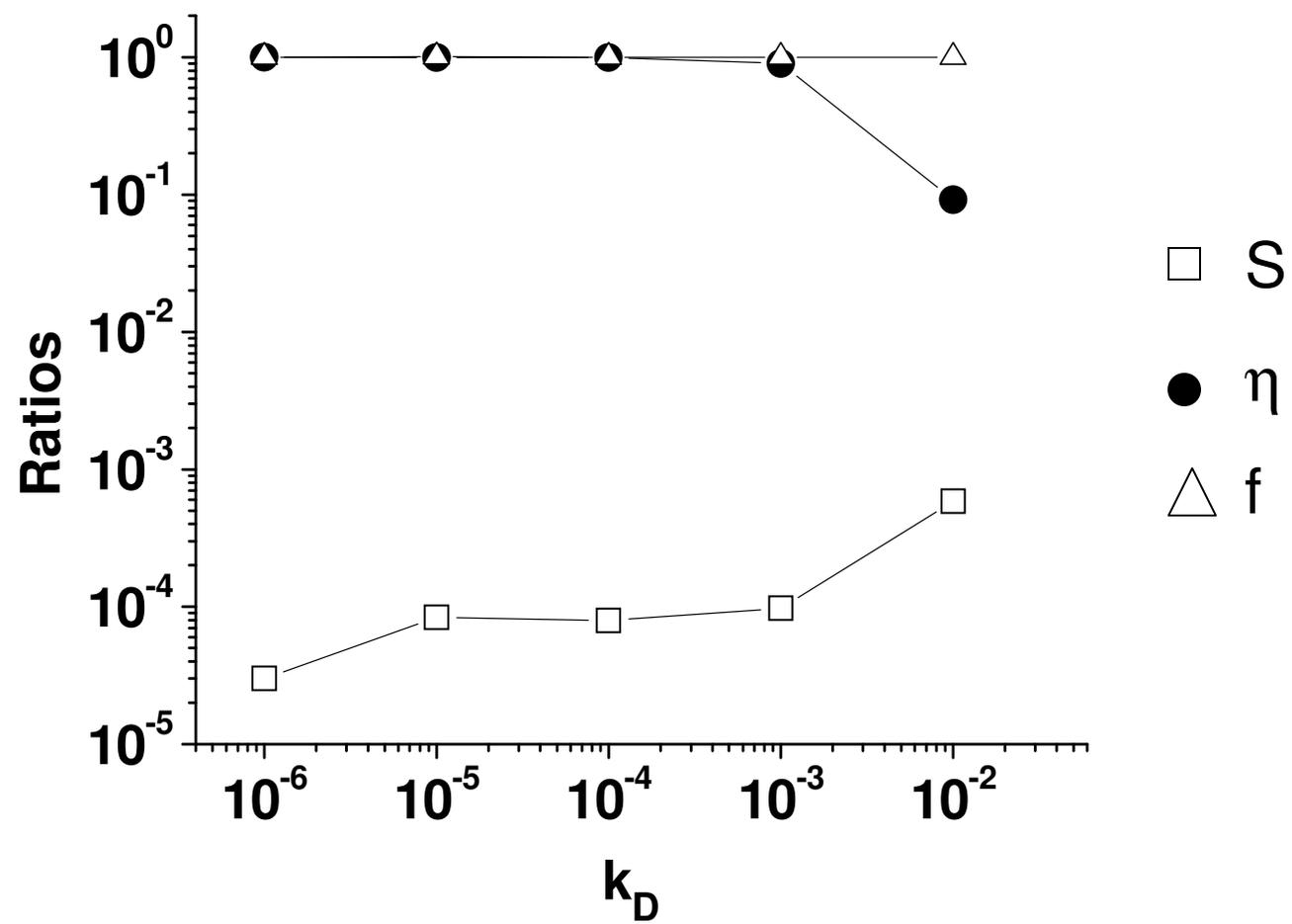

Fig 4c

Fig 4d

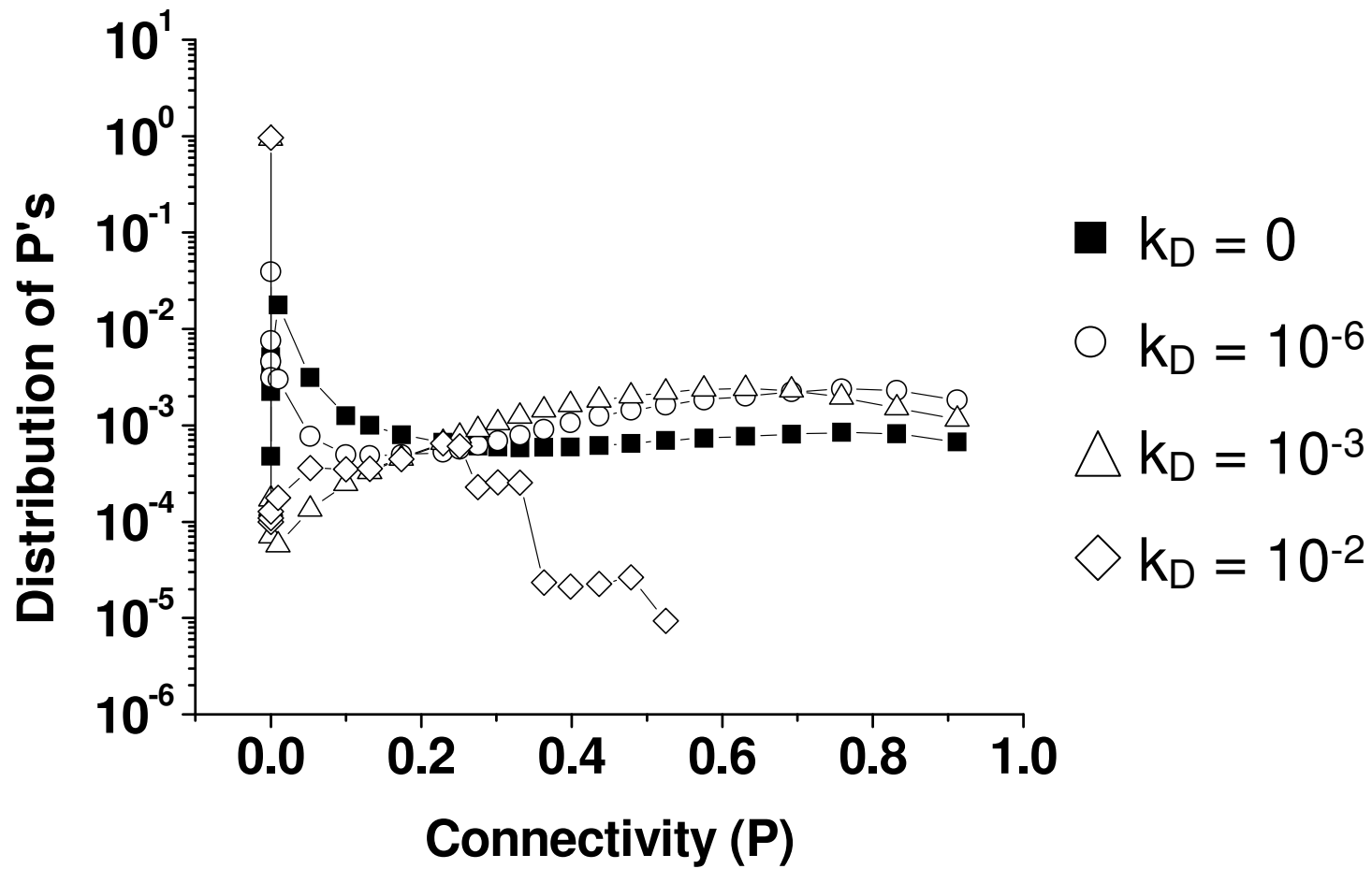

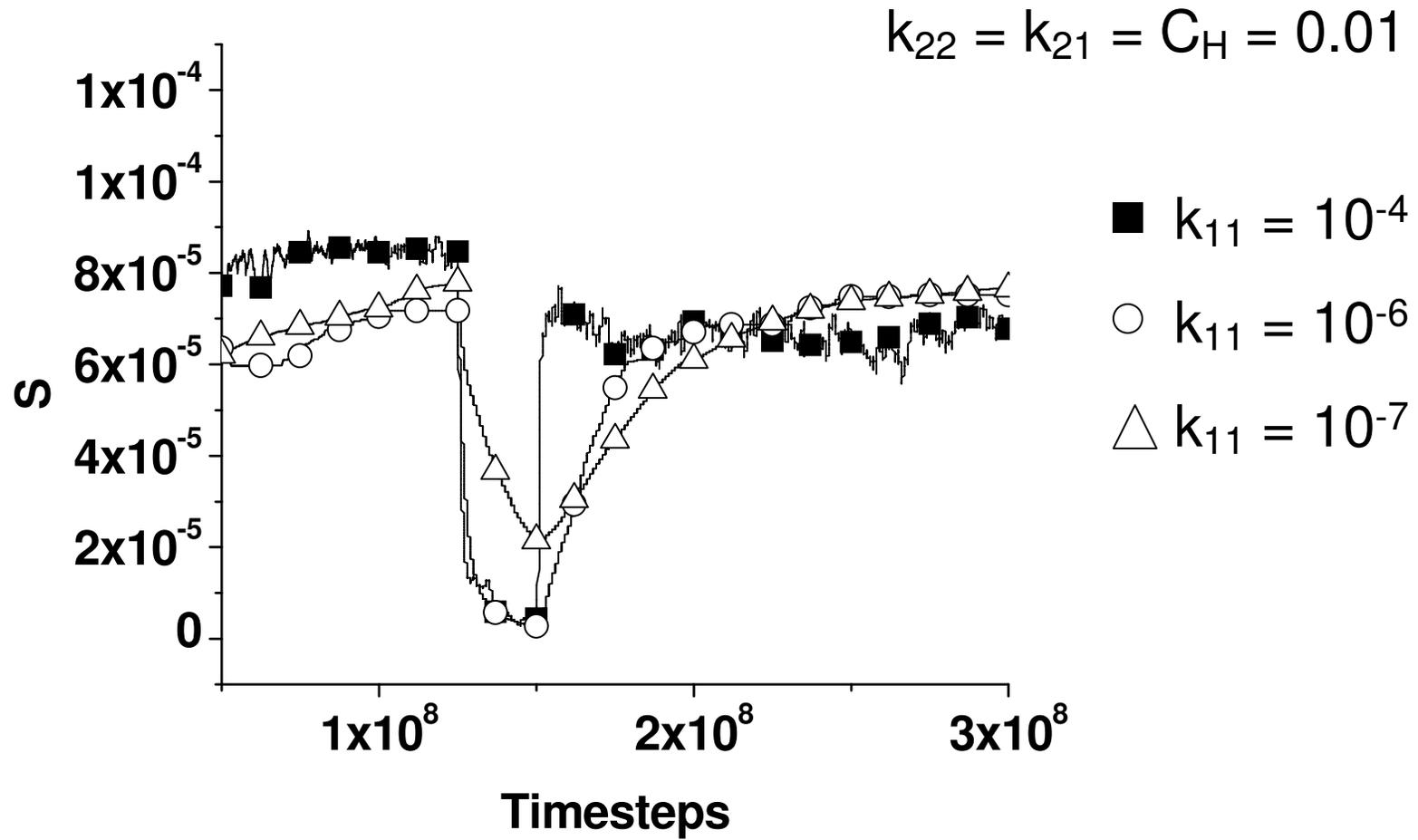

Fig 5a

Fig 5b

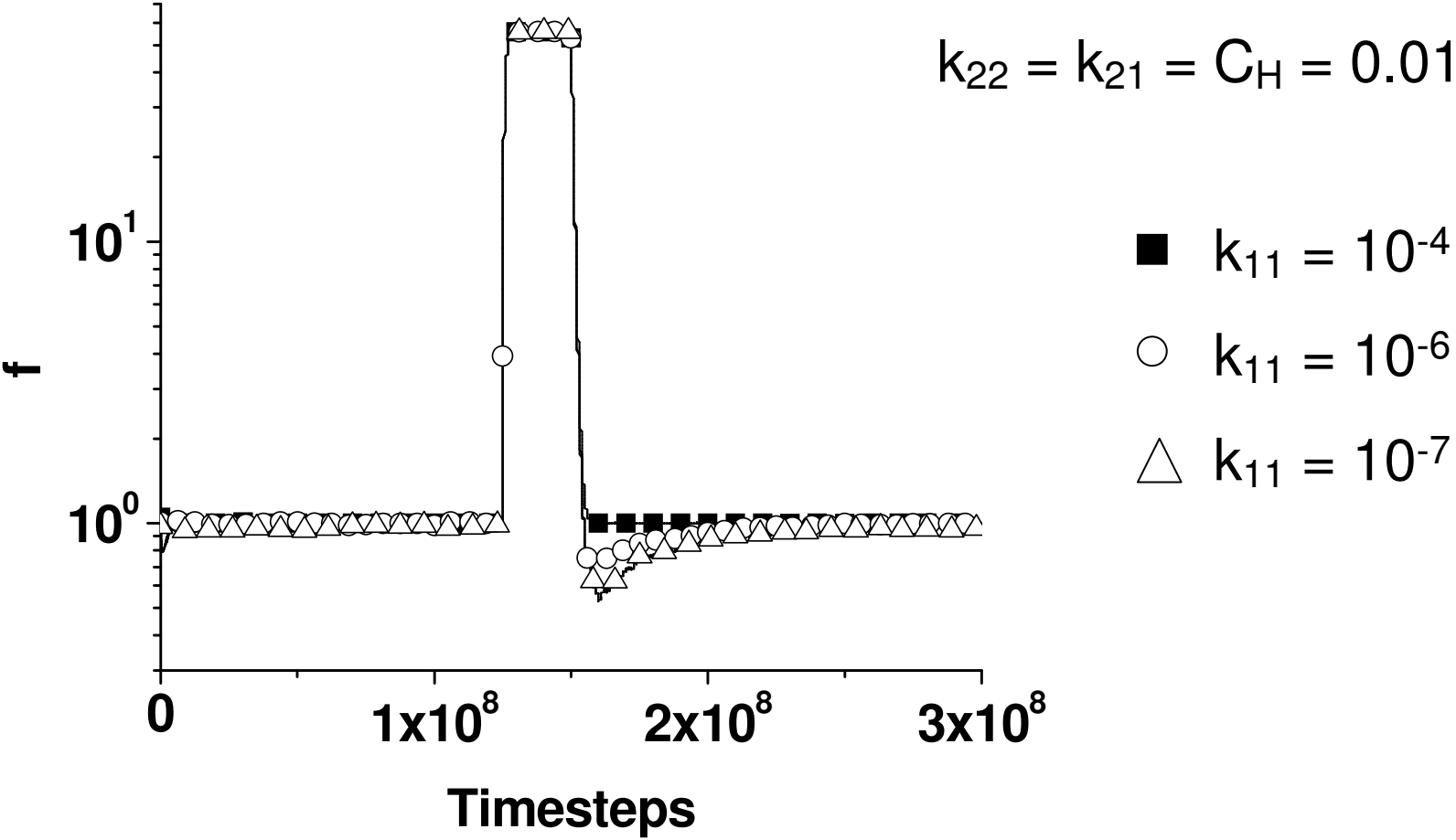

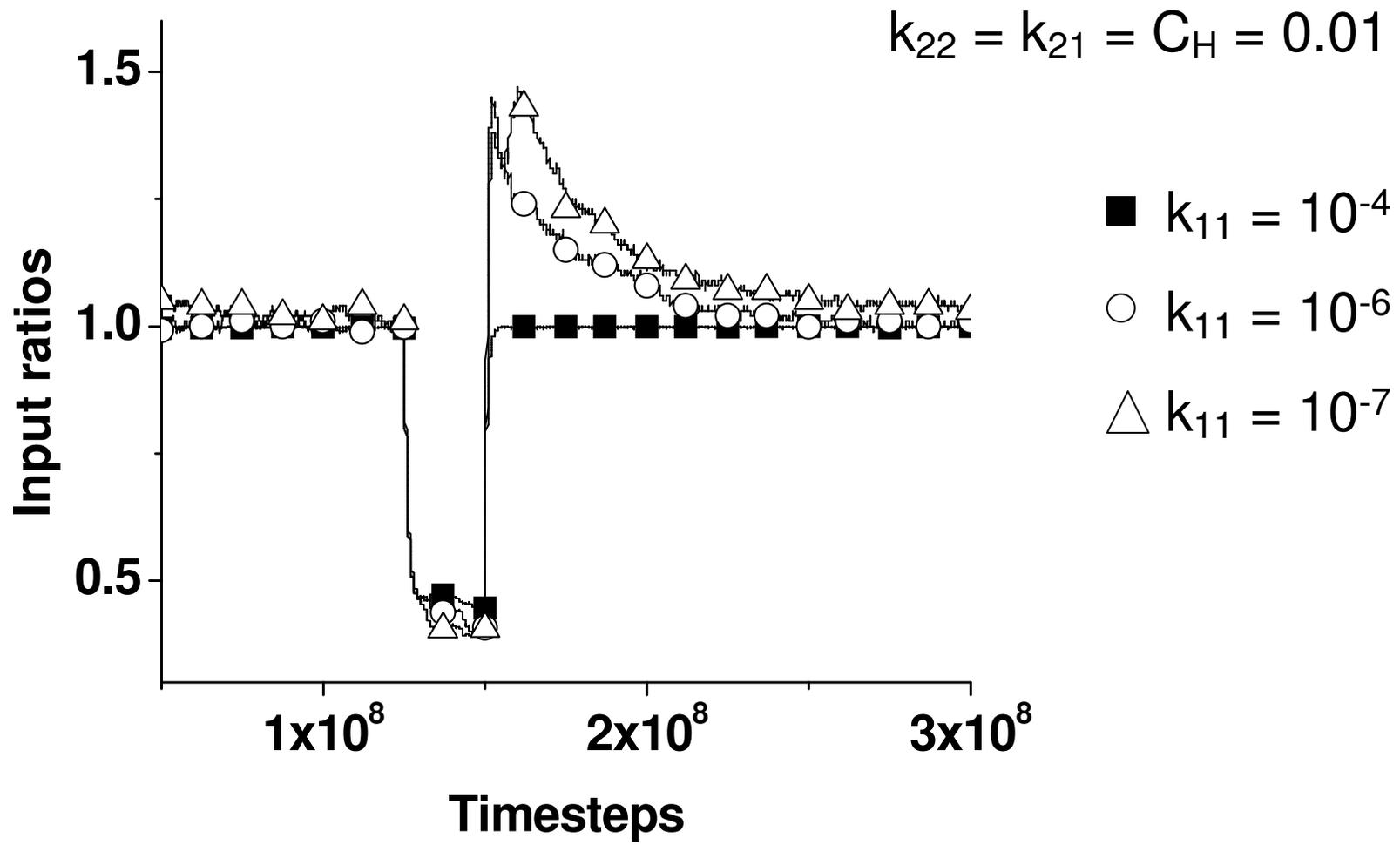

Fig 5c

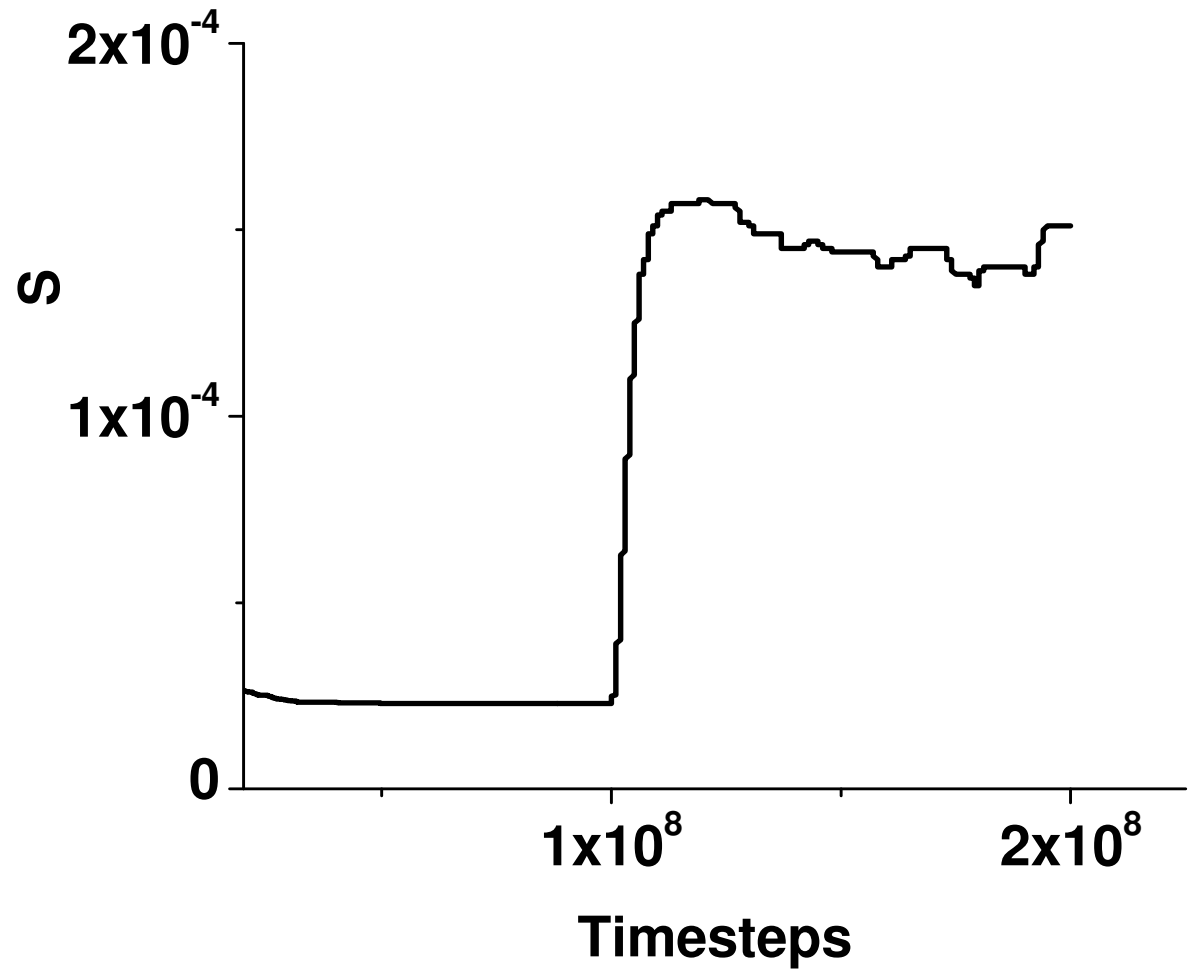

Fig 6a

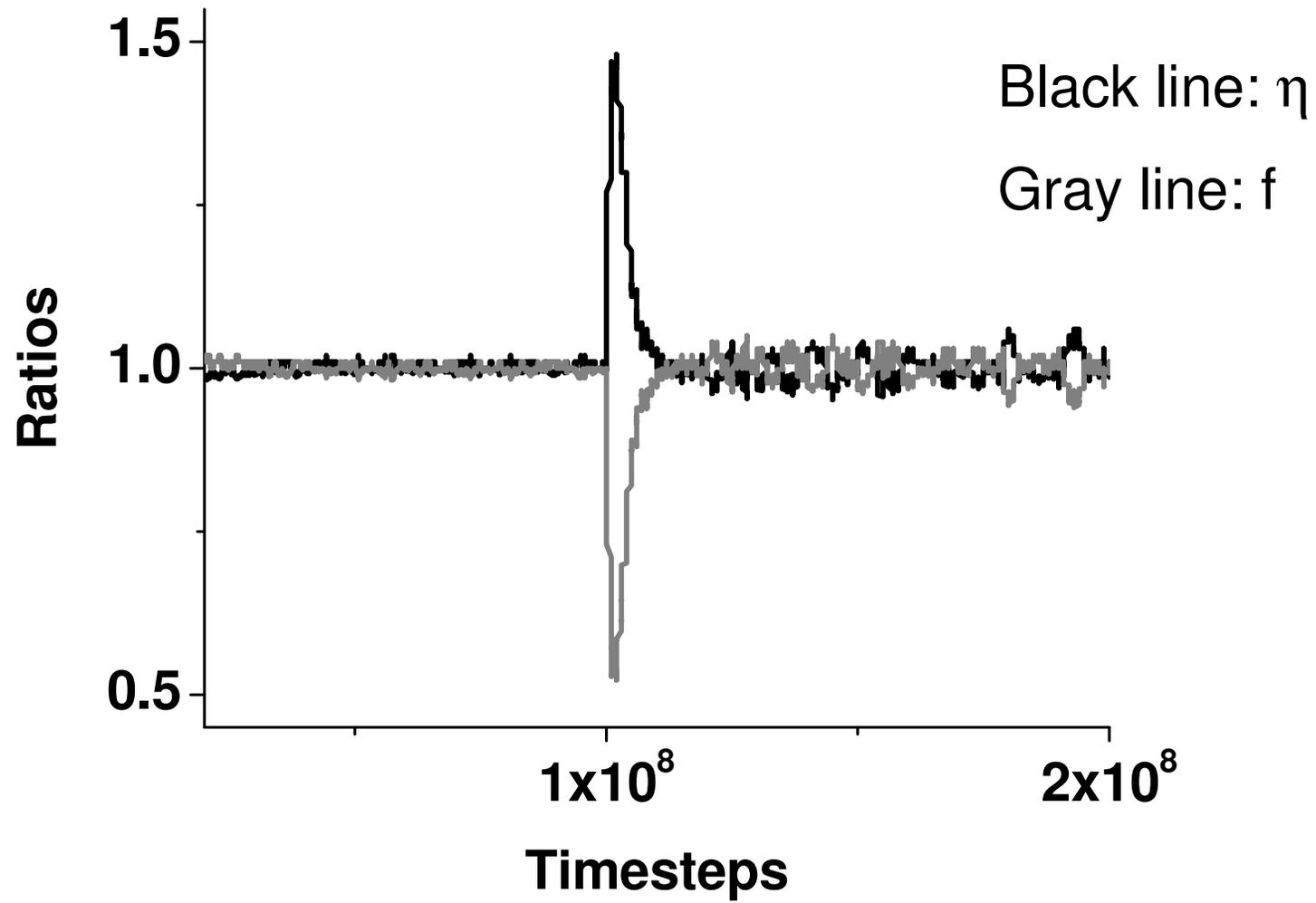

Fig 6b

Fig 7

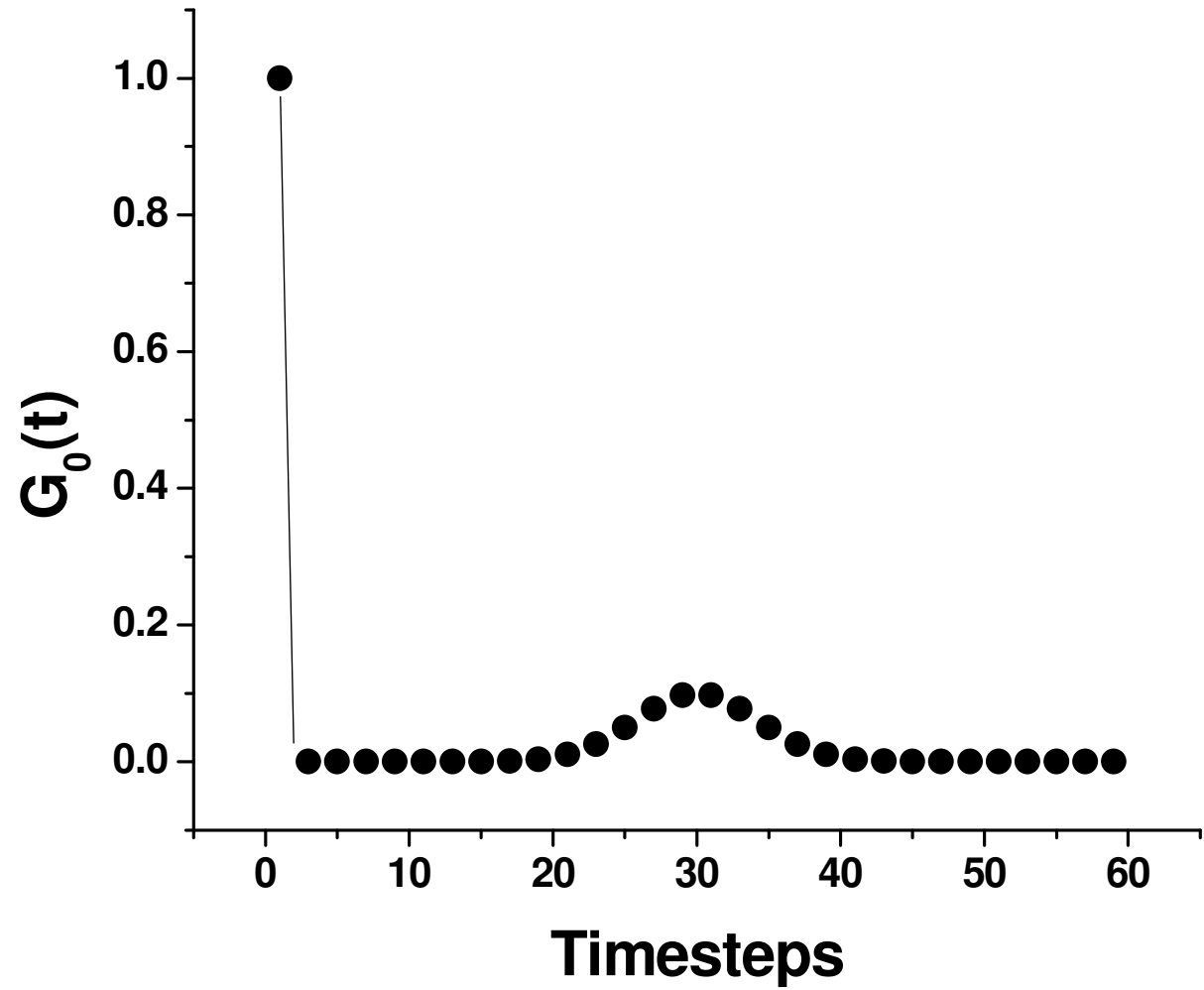